\documentclass{jfm}
\usepackage{graphicx}
\usepackage{epstopdf, epsfig}

\usepackage{mathtools}
\usepackage{amsmath}
\usepackage{amssymb}

\shorttitle{Lagrangian velocity gradient closure}
\shortauthor{P. L. Johnson and C. Meneveau}

\title{A closure for Lagrangian velocity gradient evolution in turbulence using recent deformation mapping of initially Gaussian fields}

\author{Perry L. Johnson %\aff{1}
  \corresp{\email{pjohns86@jhu.edu}},
  \and Charles Meneveau %\aff{1}
  }

\affiliation{ %\aff{1}
Department of Mechanical Engineering and Center for Environmental
and Applied Fluid Mechanics, The Johns Hopkins University,
Baltimore, MD 21218, USA
}

\begin{document}

\maketitle

\begin{abstract}
The statistics of the velocity gradient tensor in turbulent flows are of both theoretical and practical importance. The Lagrangian view provides a privileged perspective for studying the dynamics of turbulence in general, and of the velocity gradient tensor in particular. Stochastic models for the Lagrangian evolution of velocity gradients in isotropic turbulence, with closure models for the pressure Hesssian and viscous Laplacian, have been shown to reproduce important features such as non-Gaussian probability distributions, skewness and vorticity strain-rate alignments.  The Recent Fluid Deformation (RFD) closure introduced the idea of mapping an isotropic Lagrangian pressure Hessian as upstream initial condition using the fluid deformation tensor. Recent work on a Gaussian fields closure, however, has shown that even Gaussian isotropic velocity fields contain significant anisotropy for the conditional pressure Hessian tensor due to the inherent velocity-pressure couplings, and that assuming an isotropic pressure Hessian as upstream condition may not be realistic. In this paper, Gaussian isotropic field statistics are used to generate more physical upstream conditions for the recent fluid deformation mapping. In this new framework, known isotropy relations can be satisfied {\it a priori} and no DNS-tuned coefficients are necessary. A detailed comparison of results from the new model, referred to as the recent deformation of Gaussian fields (RDGF) closure, with existing models and DNS shows the improvements gained, especially in various single-time statistics of the velocity gradient tensor at moderate Reynolds numbers. Application  to arbitrarily high Reynolds numbers remains  an open challenge for this type of model, however.
 \end{abstract}

%\begin{keywords}
%Authors should not enter keywords on the manuscript, as these must be chosen by the author during the online submission process and will then be added during the typesetting process (see http://journals.cambridge.org/data/\linebreak[3]relatedlink/jfm-\linebreak[3]keywords.pdf for the full list)
%\end{keywords}

\section{Introduction}

% Motivation
The statistics of velocity gradients in isotropic turbulence are of both practical and theoretical importance in the study of turbulent flows \citep{Sreenivasan1997,Wallace2009}. The hypothesis of approximate local isotropy at sufficiently high Reynolds number \citep{Kolmogorov1941} provides an important framework for exploring universality of small-scale statistics, including velocity gradients, across a wide range of flows. It is well accepted that the dynamics of turbulence, including velocity gradients, can be better understood in a Lagrangian frame following the flow \citep{Falkovich2001, Toschi2009}. Also, in many practical situations, the velocity gradient along Lagrangian or inertial particle paths determines the dynamics of sub-Kolmogorov scale objects immersed in turbulent flows, such as the deformation and break-up of bubbles and immiscible drops \citep{Biferale2010, Maniero2012, Biferale2014}, the stretching of polymers \citep{Balkovsky2000, Chertkov2000, Bagheri2012}, the rotation rate and nutrient uptake of small swimming organisms \citep{Batchelor1980, Pedley1992, Karp-Boss1996, Parsa2012, Chevillard2013}, and hemolysis in red-blood cells \citep{Arora2004, Behbahani2009, deTullio2012}, among other applications.

% Concern of this paper
Meanwhile, from a theoretical perspective, the statistics of velocity gradients and increments in isotropic turbulence are key ingredients in exploring internal intermittency and multi-fractality \citep{Kolmogorov1962, Oboukhov1962, ParisiFrisch1985, Meneveau1991}. In recent decades, the Lagrangian view of intermittency in turbulence has become a topic of increasing interest \citep{Falkovich2001, Toschi2009}. The energy cascade has been posed in the Lagrangian frame \citep{Meneveau1994, Yu2010} and Lagrangian multi-fractality has been explored \citep{Boffetta2002, Chevillard2003, Biferale2004, Biferale2008, Arneodo2008}. While analysis methods for dynamical systems are often impractical because of the high-dimensionality of turbulent flows \citep{Frisch1995}, tools such as finite-time Lyapunov exponents are useful in the Lagrangian frame for studying chaotic advection \citep{Ottino1989} and coherent structures \citep{Haller2000, Haller2000a, Green2007, Haller2015}.

% Review of works on Lagrangian velocity gradient evolution equation
The evolution of velocity gradients along Lagrangian paths contains two unclosed terms requiring models: the deviatoric part of the pressure Hessian and the viscous Laplacian. Removal of these two terms results in the restricted Euler equation, which has the unfortunate property of leading to a finite-time singularity \citep{Vieillefosse1982,Vieillefosse1984,Cantwell1992}. The term driving the singularity is the quadratic self-amplification of velocity gradients that is kinematic in nature. The unclosed terms are evidently responsible for opposing the restricted Euler singularity. A number of studies have shed some light on the dynamics of velocity gradients along Lagrangian paths, exploring invariant spaces and the unclosed terms \citep{Nomura1998,Martin1998a,Ooi1999,Luthi2009,Lawson2015}.

% Modeling
Meanwhile, as reviewed in \cite{Meneveau2011}, attempts at closure models for the ODE governing the Lagrangian evolution of the velocity gradient tensor have enjoyed some success.  The addition of a linear relaxation term eliminates the singularity for some initial conditions, but not for all \citep{Martin1998}. \citet{Girimaji1990} introduced a model for the pressure Hessian and viscous Laplacian designed to reproduce log-normal statistics for the pseudo-dissipation by construction. \citet{Jeong2003} constructed a non-linear relaxation model for the viscous Laplacian using the trace of the inverse Cauchy-Green tensor, neglecting the deviatoric part of the pressure Hessian. \citet{Chevillard2006} and \citet{Chevillard2008} used the insight of Jeong and Girimaji's viscous Laplacian and the tetrad model of \citet{Chertkov1999} to introduce the Recent Fluid Deformation (RFD) approximation, providing closure for both the viscous Laplacian and the deviatoric part of the pressure Hessian. The underlying concept in the RFD model is that the conditional pressure Hessian can be approximated by considering an initially isotropic tensor subjected to fluid deformation for a short time using a constant velocity gradient. It was demonstrated that the RFD closure was able to prevent the singularity and the resulting system could thus reproduce many well-known velocity gradient characteristics, including trends over a small range of moderate $\Rey_\lambda$ \citep{Chevillard2006,Chevillard2008,Chevillard2011}. Increasing the Reynolds number further, however, led to unphysical results, which were studied in some detail \citep{MartinsAfonso2010}. Meanwhile, \citet{Suman2009,Suman2011} have worked on a similar modeling approach for the Lagrangian velocity gradient evolution in compressible flows .

\citet{Wilczek2014} took a different approach to closure, using a Gaussian fields (GF) assumption to compute directly the conditional averages of the deviatoric part of the pressure Hessian and viscous Laplacian in incompressible, isotropic turbulence. When using the resulting pressure Hessian from the GF model, however, \citet{Wilczek2014} found that the model was too weak to prevent the singularity. To make the model work, the Gaussian-derived coefficients were tuned empirically using stochastic estimation based on DNS data. The resulting Enhanced Gaussian Fields (EGF) closure with the fitted parameters provided good predictions of velocity gradient statistics, rivaling those of the RFD model.

The GF closure thus elucidated an important point, that even in an isotropic Gaussian velocity field, the conditional pressure Hessian tensor is not an isotropic tensor. In this paper, we propose that the initial conditions for a recent deformation closure are better represented by those of an isotropic Gaussian velocity field than by  assuming an isotropic tensor  as in the RFD closure. With this insight, we develop a new pressure Hessian and viscous Laplacian model based on a recent-deformation mapping closure  for incompressible turbulence that assumes the initial condition of the mapping to be an isotropic Gaussian velocity field.

More detailed background on the RFD and GF/EGF closures  is presented briefly in \S\ref{sec_background}. Following that, the new model based on recent deformations of Gaussian fields is introduced and explained in \S\ref{sec_new_model}. After a brief explanation in \S\ref{sec_numerical} of numerical methods for the stochastic ODEs and for the DNS data to which results are compared, an examination of results is given in \S\ref{sec_results}. The results of the new model are compared alongside RFD and EGF results with DNS data, and afterward appropriate conclusions are drawn.
 
\section{Background\label{sec_background}}

In this section, the equations for the Lagrangian evolution of the velocity gradient tensor are briefly summarized. After that, the closure approach based on the Fokker-Planck equation for the velocity gradient is reviewed. Within this paradigm, the prior RFD and GF closure models are explained.

\subsection{Lagrangian Velocity Gradient Evolution}

In this paper, we consider incompressible, Newtonian fluids with arbitrary solenoidal forcing.
%\begin{equation}
%\frac{\partial u_i}{\partial t} + u_k \frac{\partial u_i}{\partial x_k} = - \frac{\partial p}{\partial x_i} + \nu \nabla^2 u_i + f_i, \hspace{0.05\linewidth} \frac{\partial u_k}{\partial x_k} = 0.
%\label{eq_BG_NavierStokes}
%\end{equation}
The gradient of the incompressible Navier-Stokes equations gives the evolution equation for velocity gradient tensor, $A_{ij} = \tfrac{\partial u_i}{\partial x_j}$,
\begin{equation}
\frac{dA_{ij}}{dt} = - A_{ik}A_{kj} - P_{ij} + \nu \nabla^2 A_{ij} + F_{ij} %, \hspace{0.05\linewidth} A_{kk} = 0.
\label{eq_BG_LagrangianA_raw}
\end{equation}
where $\tfrac{d}{dt} = \tfrac{\partial}{\partial t} + u_k \frac{\partial}{\partial x_k}$ represents the material derivative, $P_{ij} = \frac{\partial^2 p}{\partial x_i \partial x_j}$ is the symmetric pressure Hessian tensor, and $F_{ij} = \frac{\partial f_i}{\partial x_j}$ is the trace-free gradient of the forcing. The first term on the right-hand side is the non-linear self-amplification term, which leads to the restricted Euler dynamics and a finite-time singularity \citep{Vieillefosse1982, Vieillefosse1984, Cantwell1992}. While this self-amplification term is closed, the pressure Hessian and viscous Laplacian terms are unclosed, requiring information from neighboring Lagrangian trajectories. 
%The forcing term is arbitrary and its details are likely unimportant when large-scale forcing is considered.

The incompressibility constraint, i.e. that the velocity gradient tensor should be trace-free, can be incorporated by evaluating the trace of the velocity gradient evolution equation, which yields the pressure Poisson equation, $P_{kk} = 2 Q$, where $Q \equiv - \frac{1}{2} A_{k\ell}A_{\ell k}$. 
%One role of the pressure Hessian is to enforce the trace-free condition. 
Solving the pressure Poisson equation and twice taking the gradient leads to \citep{Ohkitani1995},
\begin{equation}
P_{ij}(\mathbf{x},t) = \frac{2}{3} Q(\mathbf{x},t) \delta_{ij} + \iiint\limits_{P.V.} \frac{Q(\mathbf{x}+\mathbf{r})}{2\pi r^3} \left( \delta_{ij} - 3\frac{r_ir_j}{r^2} \right)  d\mathbf{r}.
\label{eq_BG_pressureHessianDecomp}
\end{equation}
The isotropic part of the pressure Hessian is local and closed, while the deviatoric part of the pressure Hessian, $P_{ij}^{(d)}$, is non-local and depends on the structure of the surrounding flow. Decomposition into isotropic and deviatoric parts gives
\begin{equation}
\frac{dA_{ij}}{dt} = - \left( A_{ik}A_{kj} + \frac{2}{3} Q \delta_{ij} \right) - P_{ij}^{(d)} + \nu \nabla^2 A_{ij} + F_{ij}.
\label{eq_BG_LagrangianA}
\end{equation}
This tensor equation represents 9 differential equations for the 9 components of the velocity gradient tensor,  of which 8 are independent.

The velocity gradient tensor can be written as a sum of symmetric and anti-symmetric parts, $A_{ij} = S_{ij} + \Omega_{ij}$, where $S_{ij} = \tfrac{1}{2} \left( A_{ij} + A_{ji} \right)$ is the strain-rate tensor and $\Omega_{ij} = \tfrac{1}{2} \left( A_{ij} - A_{ji} \right)$ is the rotation rate tensor, which can be related to the vorticity, $\omega_i = - \epsilon_{ijk} \Omega_{jk}$. Using this decomposition on the Lagrangian evolution equation \citep{Nomura1998},
\begin{eqnarray}
\frac{dS_{ij}}{dt} = - \left( S_{ik}S_{kj} - \frac{1}{3} S_{k\ell}S_{k\ell} \delta_{ij} \right) - \left( \Omega_{ik} \Omega_{kj} - \frac{1}{3} \Omega_{ik}\Omega_{kj} \delta_{ij}  \right) \nonumber\\  - P_{ij}^{(d)} + \nu \nabla^2 S_{ij} + F_{ij}^{(s)},
\label{eq_BG_LagrangianS}
\end{eqnarray}
\begin{equation}
\frac{d\Omega_{ij}}{dt} = - \left( S_{ik}\Omega_{kj} + \Omega_{ik}S_{kj} \right) + \nu \nabla^2 \Omega_{ij} + F_{ij}^{{(a)}},
\label{eq_BG_LagrangianW}
\end{equation}
where $F_{ij}^{(s)} = \frac{1}{2} \left( F_{ij} + F_{ji} \right)$ and $F_{ij}^{(a)} = \frac{1}{2} \left( F_{ij} - F_{ji} \right)$, are the symmetric and anti-symmetric parts of the forcing, respectively. In this way, we can separately view the evolution of the vorticity and the strain-rate, although the strong coupling in the non-linear term is evident.

\subsection{Stochastic Model}

In order to model the Lagrangian evolution of the velocity gradient, a stochastic representation is taken \citep{Girimaji1990,Chevillard2008,Wilczek2014}. The stochastic approach is built on the evolution equation for the single-time probability density function (PDF) for the velocity gradient tensor, %obtained from \eqref{eq_BG_LagrangianA},
\begin{eqnarray}
\frac{\partial f(\mathcal{A}; t)}{\partial t} & = & - \frac{\partial}{\partial \mathcal{A}_{ij}} \left(\left[ - \left( \mathcal{A}_{ik}\mathcal{A}_{kj} + \frac{2}{3} \mathcal{Q} \delta_{ij} \right) - \left\langle \left. P_{ij}^{(d)} \right| \mathcal{A} \right\rangle + \left\langle \left. \nu \nabla^2 A_{ij} \right| \mathcal{A} \right\rangle \right] f(\mathcal{A}; t)\right) \nonumber\\ 
&& + \frac{1}{2} D_{ijk\ell} \frac{\partial^2 f(\mathcal{A}; t)}{\partial \mathcal{A}_{ij} \partial \mathcal{A}_{k\ell}}.
\label{eq_BG_LagrangianApdf}
\end{eqnarray}
This Fokker-Planck equation for the PDF evolution is constructed from  \eqref{eq_BG_LagrangianA} using stochastic forcing. The related Langevin equation is,
\begin{equation}
dA_{ij} = \left[-\left( A_{ik}A_{kj} + \frac{2}{3} Q \delta_{ij} \right) - \left\langle \left. P_{ij}^{(d)} \right| \mathbf{A} \right\rangle + \left\langle \left. \nu \nabla^2 A_{ij} \right| \mathbf{A} \right\rangle \right] dt + dF_{ij},
\label{eq_BG_LagraingianStochasticA}
\end{equation}
which provides a model for the Lagrangian velocity gradient dynamics provided the two conditional averages and the stochastic noise term can be specified. The stochastic forcing term, $dF_{ij} = b_{ijk\ell} dW_{kl}$, is built on a tensorial Wiener process, $\left\langle dW_{ij} \right\rangle = 0$, $\left\langle dW_{ij} dW_{k\ell} \right\rangle = \delta_{ik} \delta_{j\ell} dt$, with diffusion tensor $D_{ijk\ell} = b_{ijmn} b_{k\ell mn}$.

Furthermore, this tensorial stochastic ODE can be decomposed into symmetric and anti-symmetric components, as in  \eqref{eq_BG_LagrangianS} and $\eqref{eq_BG_LagrangianW}$,
\begin{eqnarray}
dS_{ij} = \left[ \astrut - \left( S_{ik}S_{kj} - \frac{1}{3} S_{k\ell}S_{k\ell} \delta_{ij} \right) - \left( \Omega_{ik} \Omega_{kj} - \frac{1}{3} \Omega_{ik}\Omega_{kj} \delta_{ij}  \right) \right. \nonumber\\
- \left. \left\langle \left. P_{ij}^{(d)} \right| \mathbf{S}, \boldsymbol{\Omega} \right\rangle + \left\langle \left. \nu \nabla^2 S_{ij} \right| \mathbf{S}, \boldsymbol{\Omega} \right\rangle \astrut\right] dt + dF_{ij}^{(s)},
\label{eq_BG_LagrangianStochasticS}
\end{eqnarray}
\begin{equation}
d\Omega_{ij} = \left[ - \left( S_{ik}\Omega_{kj} + \Omega_{ik}S_{kj} \right) + \left\langle \left. \nu \nabla^2 \Omega_{ij} \right| \mathbf{S}, \boldsymbol{\Omega} \right\rangle \right] dt + dF_{ij}^{{(a)}}.
\label{eq_BG_LagrangianStochasticW}
\end{equation}
In this system, $\Omega_{ij}$ has three independent variables with the requirement to remain anti-symmetric and $S_{ij}$ has five independent variables with the requirement to remain symmetric and trace-free. The symmetric and anti-symmetric stochastic forcing terms, in this view, can be chosen independent of each other and obeying these constraints. The details of the stochastic forcing term are given in Appendix \ref{app_forcing}.

%It is worthwhile to note that this stochastic model is built on the evolution equation of the single-time PDF. Therefore, in the authors' view, it is not a-priori suitable for describing multi-time statistics, such as autocorrelation functions. Rather, it is a subject for empirical investigation as to the success of such a model in reproducing multi-time statistics of the velocity gradient tensor.

\subsection{Recent Fluid Deformation Closure}

The central idea in the RFD closure approach \citet{Chevillard2006} is to introduce a coordinate mapping based on material volume deformation in the recent Lagrangian history. Defining a Lagrangian trajectory as a map, $\mathcal{T}_{t_0,t}: \mathbf{X} \in \mathbb{R}^3 \mapsto \mathbf{x} \in \mathbb{R}^3$, from an initial condition $\mathbf{X}$ at time $t_0$ to a position $\mathbf{x}$ at a later time $t$, then the Lagrangian trajectory evolves according to $\frac{dx_i}{dt} = u_i(\mathbf{x},t)$, with initial condition $x_i(t_0) = X_i$, where the velocity field, $\mathbf{u}(\mathbf{x},t)$, is a solution to the incompressible Navier-Stokes equations with appropriate boundary and initial conditions.

The evolution of an infinitesimal fluid volume in the vicinity of $\mathbf{x}$ can be described by the deformation tensor, $D_{ij} = \partial x_i / \partial X_j$, which is the sensitivity of the trajectory to initial position and evolves as $\frac{dD_{ij}}{dt} = A_{ik} D_{kj}$ with initial condition $D_{ij}(\mathbf{X},t_0) = \delta_{ij}$. The general solution to this equation involves the time-ordered exponential, but approximating that the velocity gradient is constant for short time, % the approximate solution is 
\begin{equation}
\mathbf{D}(\mathbf{x},t; \mathbf{X}, t_0) \approx \exp\left[ \mathbf{A}(\mathbf{x},t) (t - t_0) \right].
\label{eq_BG_recentDeformationSoln}
\end{equation}
Instead of directly attempting to close the conditional averages in \eqref{eq_BG_LagraingianStochasticA}, first the approximate fluid deformation tensor is used to strain the coordinate system,
\begin{equation}
\left\langle \left. P_{ij} \right| \mathbf{A} \right\rangle = \left\langle \left. \frac{\partial^2 p}{\partial x_i \partial x_j} \right| \mathbf{A} \right\rangle \approx \frac{\partial X_k}{\partial x_i} \left\langle \left. \frac{\partial^2 p}{\partial X_k \partial X_\ell} \right| \mathbf{A} \right\rangle \frac{\partial X_\ell}{\partial x_j} = D_{ki}^{-1} \left\langle \left.  \widetilde{P}_{k\ell} \right| \mathbf{A} \right\rangle D_{\ell j}^{-1},
\label{eq_BG_mappedPressureHessian}
\end{equation}
where $\widetilde{P}_{ij}$ represents an approximation for the pressure Hessian at a previous time along the Lagrangian path and $D_{ij}^{-1} = \partial X_i / \partial x_j \approx \left(\exp\left[ - \mathbf{A}(\mathbf{x},t) (t - t_0) \right] \right)_{ij}$.   This is akin to assuming the pressure approximately constant along pathlines for a short time (in the sense of conditional averages on $\mathbf{A}$), so that the changes in the conditional pressure Hessian are due entirely to the relative movement of neighboring fluid trajectories induced by the local velocity gradient. In this way, the closure of the conditional pressure Hessian requires the specification of initial conditions of the pressure Hessian upstream along the Lagrangian path.

The strongest assumption in the RFD model comes next, in assuming the initial condition for the mapping, i.e. the upstream conditional pressure Hessian, to be an isotropic tensor,
\begin{equation}
\left\langle \left. \widetilde{P}_{ij} \right| \mathbf{A} \right\rangle \approx \frac{1}{3} \left\langle \left. \widetilde{P}_{kk} \right| \mathbf{A} \right\rangle \delta_{ij},
\label{eq_BG_isotropicPressureHessian}
\end{equation}
which gives
\begin{equation}
\left\langle \left. P_{ij} \right| \mathbf{A} \right\rangle \approx \frac{1}{3} C_{ij}^{-1} \left\langle \left. \widetilde{P}_{kk} \right| \mathbf{A} \right\rangle.
\label{eq_BG_pressureHessian1}
\end{equation}
where $C_{ij}^{-1} = D_{ki}^{-1} D_{kj}^{-1}$ is the inverse of the left Cauchy-Green tensor. The trace of  \eqref{eq_BG_pressureHessian1},
\begin{equation}
2Q = \left\langle \left. P_{kk} \right| \mathbf{A} \right\rangle \approx \frac{1}{3} C_{kk}^{-1} \left\langle \left. \widetilde{P}_{\ell\ell} \right| \mathbf{A} \right\rangle,
\label{eq_BG_pressureHessianTrace}
\end{equation}
upon substitution, allows for the final form,
\begin{equation}
\left\langle \left. P_{ij} \right| \mathbf{A} \right\rangle \approx 2Q \frac{C_{ij}^{-1}}{C_{kk}^{-1}}.
\label{eq_BG_RFD_pressureHessian}
\end{equation}
This form of the conditional pressure Hessian is appealing due to its simplicity and the intuition that the statistical bias of the pressure Hessian is related to the recent deformation of fluid particles by the velocity gradient tensor. However, as will be recalled later, even isotropic Gaussian velocity fields contain anisotropic contributions for the conditional pressure Hessian, casting some doubts onto  \eqref{eq_BG_isotropicPressureHessian} above.

The RFD model treats the viscous Laplacian in the same way,
\begin{equation}
\left\langle \left. \nu \nabla^2 A_{ij} \right| \mathbf{A} \right\rangle \approx \nu \frac{\partial X_p}{\partial x_k} \left\langle \left. \frac{\partial A_{ij}}{\partial X_p \partial X_q} \right| \mathbf{A} \right\rangle \frac{\partial X_q}{\partial x_k} = \nu D_{pk}^{-1} \left\langle \left. \frac{\partial A_{ij}}{\partial X_p \partial X_q} \right| \mathbf{A} \right\rangle D_{qk}^{-1},
\label{eq_BG_mappedViscousLaplacian}
\end{equation}
and assumes that the conditional Hessian of the velocity gradient tensor is likewise an isotropic tensor,
\begin{equation}
\left\langle \left. \frac{\partial A_{ij}}{\partial X_k \partial X_\ell} \right| \mathbf{A} \right\rangle = \frac{1}{3} \left\langle \left. \nabla_{\mathbf{X}}^2 A_{ij} \right| \mathbf{A} \right\rangle \delta_{k\ell},
\label{eq_BG_isotropicViscousLaplacian}
\end{equation}
which yields,
\begin{equation}
\left\langle \left. \nu \nabla^2 A_{ij} \right| \mathbf{A} \right\rangle \approx \frac{C_{kk}^{-1}}{3}  \nu \left\langle \left. \nabla_{\mathbf{X}}^2 A_{ij} \right| \mathbf{A} \right\rangle.
\label{eq_BG_ViscousLaplacian1}
\end{equation}
Then taking a linear relaxation model \citep{Martin1998} for the initial upstream conditions of the conditional viscous Laplacian,
\begin{equation}
\left\langle \left. \nu \nabla^2 A_{ij} \right| \mathbf{A} \right\rangle \approx - \frac{C_{kk}^{-1}}{3} \frac{A_{ij}}{T}.
\label{eq_BG_RFD_ViscousLaplacian}
\end{equation}
In this way, the recent deformation provides a physically motivated mechanistic approach to introduce non-linearity in the viscous Laplacian term, which helpful in removing the finite-time singularity (the Cauchy-Green tensor is exponential rather than linear). \citet{Chevillard2006} and \citet{Chevillard2008} argue that the proper relaxation timescale, $T$, for the viscous Laplacian is the integral timescale, and that the proper timescale for the recent deformation is the Kolmogorov timescale, $t - t_0 = \tau_\eta$. In this way, the model introduces $\Rey_\lambda \sim  ({T}/{\tau_\eta})$ effects. Indeed, certain intermittency trends are reproduced by this model \citep{Chevillard2006} at moderate $\Rey_\lambda$, but continuing to increase $\Rey_\lambda$ beyond a certain threshold leads to increasingly unphysical results \citep{MartinsAfonso2010}. Nonetheless, the RFD closure provides a model that reproduced many of the known trends of velocity gradient statistics at moderate $\Rey_\lambda$, and continues to be useful for studying velocity gradient statistics \citep{Moriconi2014}.

\subsection{Gaussian Fields Closure}

\citet{Wilczek2014} took a different approach to closing the conditional averages. They assumed that the {\it velocity} field has joint-Gaussian N-point PDFs with prescribed spectral (two-point) statistics (the pressure field constructed from such a velocity field is not necessarily Gaussian). They computed the conditional averages using this approximation by employing the characteristic functional of a Gaussian velocity field and obtaining an exact analytical result for the conditional pressure Hessian for a Gaussian velocity field 
\begin{eqnarray}
\left\langle \left. P_{ij}^{(d)} \right| \mathbf{A} \right\rangle_{\text{Gaussian}} &=& \alpha \left( S_{ik}S_{kj} - \frac{1}{3} S_{k\ell}S_{\ell k} \delta_{ij} \right) + \beta \left( \Omega_{ik}\Omega_{kj} - \frac{1}{3} \Omega_{k\ell}\Omega_{\ell k} \delta_{ij} \right) \nonumber\\
&& + \gamma \left( S_{ik}\Omega_{kj} - \Omega_{ik}S_{kj} \right)
\label{eq_BG_GaussianPressureHessian}
\end{eqnarray}
where
\begin{equation}
\alpha = -\frac{2}{7}, \hspace{0.05\linewidth} \beta = -\frac{2}{5}, \hspace{0.05\linewidth} \gamma = \frac{6}{25} + \frac{16}{75 f''(0)^2} \int \frac{f'(r)f'''(r)}{r} dr,
\label{eq_BG_GaussianCoeffs}
\end{equation}
with $f(r)$ specifying the longitudinal velocity correlation function in isotropic turbulence. In this expression, $\alpha$ and $\beta$ are independent of $\Rey_\lambda$ while $\gamma$ is expected to have weak $\Rey_\lambda$-dependence through the integral of the correlation function derivatives. Using a model spectrum at $\Rey_\lambda = 430$, a numerical result of $\gamma \approx 0.08$ was obtained \citep{Wilczek2014} .

Furthermore, the conditional viscous Laplacian could  also be computed for Gaussian fields \citep{Wilczek2014},
\begin{equation}
\left\langle \left. \nu \nabla^2 A_{ij} \right| \mathbf{A} \right\rangle_{\text{Gaussian}} = \delta A_{ij}, ~~~{\rm where} ~~~ \delta = \nu \frac{7}{3} \frac{f^{(4)}(0)}{f''(0)}.
%\label{eq_BG_GaussianCoeff}
\label{eq_BG_GaussianViscousLaplacian}
\end{equation}
Note that $\delta < 0$ for realistic correlation functions, meaning that the Gaussian approximation leads to a linear damping model as in \citet{Martin1998} for the viscous Laplacian. Numerical evaluation using a model spectrum at $\Rey_\lambda = 430$ gave the result $\delta \approx -0.65 / \tau_\eta$. Using the above Gaussian-derived functional form but invoking in addition the balance of enstrophy production and dissipation with its relationship to skewness,   \citet{Wilczek2014}  related the coefficient $\gamma$ to the velocity derivative skewness, $\mathcal{S} = \left\langle \left( \partial u_1 / \partial x_1 \right)^3 \right\rangle / \left\langle \left( \partial u_1 / \partial x_1 \right)^2 \right\rangle^{3/2}$, 
\begin{equation}
\delta = \frac{7}{6 \sqrt{15}} \frac{\mathcal{S}}{\tau_\eta},
\label{eq_BG_GaussianCoeffSkewness}
\end{equation}
a result which gave much better agreement values estimated from DNS at $\Rey_\lambda = 430$, namely $\delta \approx -0.15 / \tau_\eta$, when using realistic values for the skewness (non-zero, i.e. non-Gaussian). Because the original Gaussian closure led to a singularity when integrated numerically, the authors considered an alternative model in which the functional form of the Gaussian closure was retained by the coefficients were empirically obtained by estimating them from DNS results:
$\alpha = -0.61,~ \beta = -0.65, ~ \gamma = 0.14, ~ \delta = -0.15 / \tau_\eta.$
%\label{eq_BG_EnhancedGaussianCoeffs}
%
With these empirically-adjusted coefficients, statistical stationarity was achieved and many of the known trends for velocity gradient statistics were reproduced with this approach termed the Enhanced Gaussian Fields (EGF) closure.

\section{Recent Deformation of Gaussian Fields Mapping Closure\label{sec_new_model}}
This section introduces the RDGF closure for the pressure Hessian and viscous Laplacian terms in the Lagrangian stochastic evolution equation for the velocity gradient tensor.

\subsection{Overview}
\begin{figure} %[htbp]
\begin{center}
\includegraphics{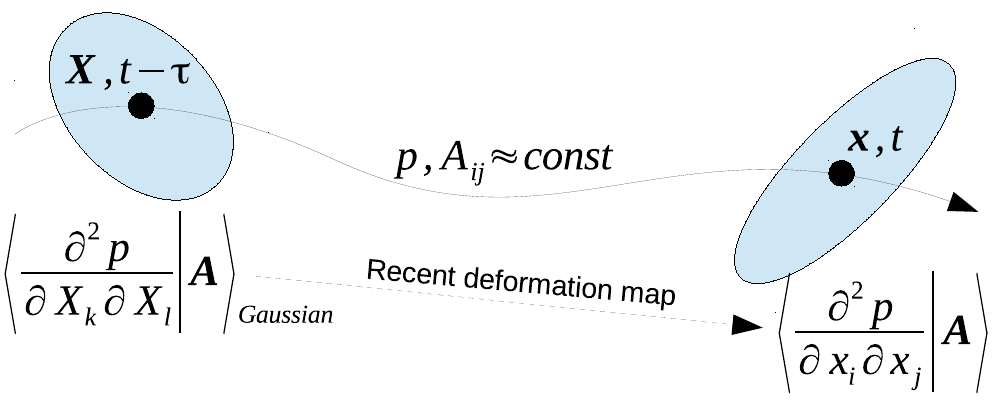}
\end{center}
\caption{Schematic illustrating the main elements of the RDGF model for the conditional pressure Hessian. The viscous Laplacian model is constructed analogously.}
\label{fig_MODEL_framework}
\end{figure}

As summarized before, a strong assumption underlying the RFD approximation was the assumption that the initial upstream condition of the conditional pressure Hessian (and viscous Laplacian) are  isotropic tensors. Here we relax this strong assumption and instead assume that the upstream conditional pressure Hessian is that of an isotropic Gaussian velocity field. In this way,  \eqref{eq_BG_isotropicPressureHessian} is modified as follows
\begin{equation}
\left\langle \left. \widetilde{P}_{ij} \right| \mathbf{A} \right\rangle \approx \frac{1}{3} \left\langle \left. \widetilde{P}_{kk} \right| \mathbf{A} \right\rangle \delta_{ij} + \left\langle \left. \widetilde{P}_{ij}^{(d)} \right| \mathbf{A} \right\rangle_{\text{Gaussian}},
\label{eq_MODEL_pressureHessian}
\end{equation}
where the latter term is evaluated using  \eqref{eq_BG_GaussianPressureHessian}. Similarly for the viscous term, the conditional Hessian of the upstream velocity gradient is no longer assumed isotropic, and  \eqref{eq_BG_isotropicViscousLaplacian} is modified to include the anisotropic contributions from the Gaussian closure.
The same mapping as in the RFD model is applied to convert the upstream initial conditions to the resulting closure. Figure \ref{fig_MODEL_framework} illustrates the overall procedure for constructing the model for the pressure Hessian. A similar procedure is used for the viscous Laplacian.

We name this approach the Recent Deformation of Gaussian Fields (RDGF) model. In the sense of this nomenclature, the term `Gaussian fields' is used to refer to the Gaussian velocity field along with its associated (non-Gaussian) pressure field. For the pressure Hessian, the recent deformation mapping is applied to the pressure field derived from Gaussian velocity field.

The underlying phenomenology of the RDGF model is that approximate turbulence statistics can be developed efficiently by a mapping of Gaussian statistics. This motivation is similar to the ideas behind the mapping closures \citep{Chen1989,Kraichnan1990,Pope1991}, as well as the multiscale turnover Lagrangian map (MTLM) procedure of \citet{Rosales2008} to generate non-Gaussian synthetic turbulence fields.

\subsection{Model Details}

The model for the unclosed terms along the Lagrangian path at point $\mathbf{x}$ (time $t$) involves applying the Gaussian fields approximation at the upstream point $\mathbf{X}$ (time $t - \tau$). For the deviatoric part of the pressure Hessian, using \eqref{eq_BG_GaussianPressureHessian},
\begin{eqnarray}
\left\langle \left. \widetilde{P}_{ij}^{(d)} \right| \mathbf{A} \right\rangle_{\text{Gaussian}} &=& \alpha \left( S_{ik}S_{kj} - \frac{1}{3} S_{k\ell}S_{\ell k} \delta_{ij} \right) + \beta \left( \Omega_{ik}\Omega_{kj} - \frac{1}{3} \Omega_{k\ell}\Omega_{\ell k} \delta_{ij} \right) \nonumber\\
&& + \gamma \left( S_{ik}\Omega_{kj} - \Omega_{ik}S_{kj} \right),
\label{eq_MODEL_GaussianPressureHessian}
\end{eqnarray}
where \eqref{eq_BG_GaussianCoeffs} provides the numerical values of the parameters for Gaussian fields. 
%\begin{equation}
%\alpha = -\frac{2}{7}, \hspace{0.05\linewidth} \beta = -\frac{2}{5}, \hspace{0.05\linewidth} \gamma = \frac{6}{25} + \frac{16}{75 f''(0)^2} \int \frac{f'(r)f'''(r)}{r} dr.
%\label{eq_MODEL_GaussianCoeffs}
%\end{equation}
In Appendix \ref{app_gamma}, an analytical evaluation of $\gamma$ using Batchelor interpolation for the second-order structure function is presented \citep{Batchelor1951}. The result, $\gamma = \frac{86}{1365} \approx 0.063$, does not deviate much from the previous numerical result \citep{Wilczek2014}.

Similarly, the Gaussian fields approximation for the upstream Hessian of the velocity gradient uses the results of Appendix \ref{app_gaussian},
\begin{eqnarray}
\left\langle \left. \nu \frac{\partial^2 A_{ij}}{\partial X_p \partial X_q} \right| \mathbf{A} \right\rangle_{\text{Gaussian}} &=& \delta \left[\astrut T_{ij} \delta_{pq} + T_{iq} \delta_{jp} + T_{ip} \delta_{jq} \right. \nonumber \\
&& \left. - \frac{2}{21} \left( S_{jq} \delta_{ip} + S_{jp} \delta_{iq} + S_{pq} \delta_{ij} \right) \astrut\right],
\label{eq_MODEL_GaussianViscousHessian}
\end{eqnarray}
where
\begin{equation}
\delta = \nu \frac{7}{3} \frac{f^{(4)}(0)}{f''(0)}, \hspace{0.05\linewidth} T_{ij} = \frac{23}{105} A_{ij} + \frac{2}{105} A_{ji}, \hspace{0.05\linewidth} S_{ij} = \frac{1}{2} \left(A_{ij} + A_{ji}\right).
\label{eq_MODEL_GaussianViscousHessian_tensors}
\end{equation}
It can be easily shown that contraction with $\delta_{pq}$ recovers  \eqref{eq_BG_GaussianViscousLaplacian} and contraction with $\delta_{ij}$, $\delta_{ip}$, or $\delta_{iq}$ causes the term to vanish in accordance with incompressibility. Following \citet{Wilczek2014}, i.e.  \eqref{eq_BG_GaussianCoeffSkewness}, the enstrophy balance is used to determine $\delta$ in Appendix \ref{app_delta},
\begin{equation}
\delta = \frac{C_{kk}}{3} \frac{7}{6\sqrt{15}} \frac{\mathcal{S}}{\tau_\eta},
\end{equation}
where the typical value of $\mathcal{S} = -0.6$ can be used.

Then, the conditional pressure Hessian and velocity gradient Hessian are mapped from $\mathbf{X}$ to $\mathbf{x}$ along the trajectory. The deformation tensor used for the mapping, $D_{ij} = \frac{\partial x_i}{\partial X_j}$, is approximated by assuming that the velocity gradient is constant for the short time span $\tau$, i.e.  \eqref{eq_BG_recentDeformationSoln}. Using \eqref{eq_BG_mappedPressureHessian} with the new upstream conditional pressure Hessian in \eqref{eq_MODEL_pressureHessian},
\begin{equation}
\left\langle \left. P_{ij} \right| \mathbf{A} \right\rangle = \frac{1}{3} \left\langle \left. \widetilde{P}_{\ell\ell} \right| \mathbf{A} \right\rangle C_{ij}^{-1} + D_{mi}^{-1} \left\langle \left. \widetilde{P}_{mn}^{(d)} \right| \mathbf{A} \right\rangle_{\text{Gaussian}} D_{nj}^{-1},
\label{eq_MODEL_mappedGaussianPressureHessian}
\end{equation}
where  \eqref{eq_MODEL_GaussianPressureHessian} is substituted for the deviatoric part of the pressure Hesssian. The trace of this equation gives,
\begin{equation}
2Q = \left\langle \left. P_{kk} \right| \mathbf{A} \right\rangle = D_{mk}^{-1} \left\langle \left. \widetilde{P}_{mn}^{(d)} \right| \mathbf{A} \right\rangle D_{nk}^{-1} + \frac{1}{3} \left\langle \left. \widetilde{P}_{\ell\ell} \right| \mathbf{A} \right\rangle C_{kk}^{-1}.
\label{eq_MODEL_traceMappedGaussianPressureHessian}
\end{equation}
Solving  \eqref{eq_MODEL_traceMappedGaussianPressureHessian} for $\left\langle \left. \widetilde{P}_{kk} \right| \mathbf{A} \right\rangle$,
%\begin{equation}
%\frac{1}{3} \left\langle \left. P_{kk} \right| \mathbf{A} \right\rangle = \frac{1}{C_{kk}^{-1}}\left( 2Q - D_{m\ell}^{-1} \left\langle \left. \widetilde{P}_{mn}^{(d)} \right| \mathbf{A} \right\rangle D_{n\ell}^{-1} \right),
%\end{equation}
and substituting into  \eqref{eq_MODEL_mappedGaussianPressureHessian}, the resulting closure is,
\begin{equation}
\left\langle \left. P_{ij} \right| \mathbf{A} \right\rangle =  2Q \frac{C_{ij}^{-1}}{C_{kk}^{-1}} + G_{ij} - \frac{C_{ij}^{-1}}{C_{kk}^{-1}} G_{\ell\ell},
\end{equation}
where 
\begin{equation}
G_{ij} = D_{mi}^{-1} \left\langle \left. \widetilde{P}_{mn}^{(d)} \right| \mathbf{A} \right\rangle_{\text{Gaussian}} D_{nj}^{-1},
\end{equation}
using  \eqref{eq_MODEL_GaussianPressureHessian} with  \eqref{eq_BG_GaussianCoeffs}. Similarly for the viscous Laplacian, using  \eqref{eq_BG_mappedViscousLaplacian} with the new upstream conditional viscous Hessian \eqref{eq_MODEL_GaussianViscousHessian} leads to,
\begin{equation}
\left\langle \left. \nu \nabla^2 A_{ij} \right| \mathbf{A} \right\rangle =
\delta \left( T_{ij} C_{kk}^{-1}  + 2 T_{ik} B_{kj}^{-1} - \frac{4}{21} B_{ik}^{-1} S_{kj} - \frac{2}{21} B_{k\ell}^{-1} S_{k\ell} \delta_{ij} \right)
\label{eq_MODEL_mappedViscousLaplacian}
\end{equation}
where $B_{ij}^{-1} = D_{ik}^{-1} D_{jk}^{-1}$ is the inverse of the right Cauchy-Green tensor and $\mathbf{T}$ and $\mathbf{S}$ are given in  \eqref{eq_MODEL_GaussianViscousHessian_tensors}.

\subsection{The Resulting Model}

The resulting stochastic ODE model for the Lagrangian velocity gradient dynamics is
\begin{equation}
dA_{ij} = \left[-\left( A_{ik}A_{kj} - \frac{C_{ij}^{-1}}{C_{kk}^{-1}} \text{tr}\left(\mathbf{A}^2\right) \right) - \left( G_{ij} - \frac{C_{ij}^{-1}}{C_{kk}^{-1}} \text{tr}\left(\mathbf{G}\right) \right) + V_{ij} \right] dt + b_{ijk\ell} dW_{ij},
\end{equation}
where the contribution of the deviatoric part of the back-in-time pressure Hessian is,
\begin{eqnarray}
G_{ij} &=& D_{mi}^{-1} \left[\astrut -\frac{2}{7} \left( S_{mk}S_{kn} - \frac{1}{3} S_{k\ell}S_{\ell k} \delta_{mn} \right) - \frac{2}{5} \left( \Omega_{mk}\Omega_{kn} - \frac{1}{3} \Omega_{k\ell}\Omega_{\ell k} \delta_{mn} \right) \right.\nonumber \\ 
&& \left. + \frac{86}{1365} \left( S_{mk}\Omega_{kn} - \Omega_{mk}S_{kn} \right) \astrut\right] D_{nj}^{-1},
\end{eqnarray}
and the contribution of the viscous Laplacian is,
\begin{equation}
V_{ij} = \frac{7}{6\sqrt{15}} \frac{C_{kk}}{3}\frac{\mathcal{S}}{\tau_\eta} \left( T_{ij} C_{kk}^{-1}  + 2 T_{ik} B_{kj}^{-1} - \frac{4}{21} B_{ik}^{-1} S_{kj} - \frac{2}{21} B_{k\ell}^{-1} S_{k\ell} \delta_{ij} \right),
\end{equation}
with $\mathcal{S} = -0.6$ and,
\begin{equation}
S_{ij} = \frac{1}{2} \left( A_{ij} + A_{ji} \right), \hspace{0.05\linewidth} \Omega_{ij} = \frac{1}{2} \left( A_{ij} - A_{ji} \right), \hspace{0.05\linewidth} T_{ij} = \frac{23}{105} A_{ij} + \frac{2}{105} A_{ji}.
\end{equation}
The recent deformation is described by
\begin{equation}
D_{ij}^{-1} = \left[\exp\left( - \mathbf{A} \tau \right)\right]_{ij}, \hspace{0.05\linewidth} C_{ij}^{-1} = D_{ki}^{-1} D_{kj}^{-1}, \hspace{0.05\linewidth} B_{ij}^{-1} = D_{ik}^{-1} D_{jk}^{-1},
\end{equation}
and the diffusion coefficient tensor of the stochastic forcing term is
\begin{equation}
b_{ijk\ell} = -\frac{1}{3} \sqrt{\frac{D_s}{5}} \delta_{ij} \delta_{k\ell} + \frac{1}{2} \left( \sqrt{\frac{D_s}{5}} + \sqrt{\frac{D_a}{3}} \right) \delta_{ik} \delta_{j\ell} + \frac{1}{2} \left( \sqrt{\frac{D_s}{5}} - \sqrt{\frac{D_a}{3}} \right) \delta_{i\ell} \delta_{jk}.
\end{equation}
Note that the present model does {\it not} use the coefficients estimated from DNS. Instead, the coefficients are used as derived from the Gaussian field statistics.

In some sense, this model can be seen as a generalization of both RFD and GF closures. To recover the RFD model, first the back-in-time deviatoric component of the pressure Hessian should be removed, $G_{ij} = 0$, i.e. $\alpha = \beta = \gamma = 0$. Then, including only the isotropic part of  \eqref{eq_MODEL_GaussianViscousHessian} gives $\nu \nabla^2 A_{ij} = \delta \tfrac{C_{kk}^{-1}}{3} A_{ij}$, and the coefficient should be set to $\delta = -\tfrac{1}{T}$, where $T$ is the integral timescale. This roughly corresponds to the RFD model at $\frac{\tau_K}{T} = -\frac{7 \mathcal{S}}{6\sqrt{15}} \approx 0.18$. To recover the GF model, the deformation tensor should be set to identity, $D_{ij} = \delta_{ij}$.

\subsection{Parameters and Constraints}

The model now contains three parameters that have yet to be determined: $D_s$, $D_a$, and $\tau$. As discussed in more detail in Appendix \ref{app_forcing}, the stochastic forcing term, $dF_{ij} = b_{ijk\ell} dW_{k\ell}$, can be split into symmetric and anti-symmetric parts, each with its own amplitude. This can be thought of as separately forcing Eqs \eqref{eq_BG_LagrangianStochasticS} and \eqref{eq_BG_LagrangianStochasticW}. The amplitudes of the symmetric and anti-symmetric stochastic forcing tensors, $D_s$ and $D_a$, are two parameters that must be set to fully specify the model.

Additionally, the time interval of the mapping, $\tau$, must be set. In keeping with the RFD phenomenology, it is expected that this should be $\tau \sim \mathcal{O}(\tau_\eta)$. The RFD model used $\tau = \tau_K$, where $\tau_K$ is an {\it input} Kolmogorov timescale, but {\it a posteriori} evaluation at $\tfrac{\tau_K}{T} = 0.1$ reveals that the actual Kolmogorov timescale produced by the model is $\tau_\eta \approx 2.0 \tau_K$. Therefore, the effective time interval was $\tau \approx 0.5 \tau_\eta$, based on the actual velocity gradient statistics produced by the model.

The three free parameters can be set by a choice of three constraints. First, without loss of generality, considering the evolution of the dimensionless velocity gradient tensor, $\left\langle S_{ij} S_{ij} \right\rangle = \frac{1}{2}$. This constraint effectively guarantees that the definition of $\delta$ in terms of $\tau_\eta$ is consistent. For the other two constraints, it is desirable to pick relationships for isotropic turbulence with analytical derivation, which can be considered {\it a priori} constraints. It is natural, then, to pick the two Betchov relations \citep{Betchov1956}, $\left\langle Q \right\rangle = \left\langle R \right\rangle = 0$. In light of the aforementioned dimensionless form of the equation, the former can be rephrased as $\left\langle \Omega_{ij} \Omega_{ij} \right\rangle = \frac{1}{2}$.

The determination of the three parameters using the three constraints can be posed as a three-dimensional root-finding problem. The appropriate values for the parameters were found empirically by numerical solutions of the model (see \S \ref{sec_SDE_numerical} below for details) using Broyden's method \citep{Press1992}. The procedure involved iteratively adjusting $D_{s}$, $D_{a}$, and $\tau$ and evaluating sufficiently converged statistics of $\left\langle S_{ij} S_{ij} \right\rangle$, $\left\langle \Omega_{ij} \Omega_{ij} \right\rangle$ and $\langle R \rangle$ from the numerical solutions of the model until the constraints were satisfied with the desired accuracy (four decimal places). The iterative method for determining the correct model parameters converges toward,
\begin{equation}
D_s = 0.1014 / \tau_\eta^3, \hspace{0.05\linewidth} D_a = 0.0505 / \tau_\eta^3, \hspace{0.05\linewidth} \tau = 0.1302 \tau_\eta.
\end{equation}
The mapping time is considerably shorter than that of RFD closure because the additional deviatoric part of the pressure Hessian was added to the RFD model, which was by itself already strong enough to counter the singularity with $\tau \approx 0.5 \tau_\eta$.

%The stochastic model of Girimaji and Pope \citep{Girimaji1990} also tuned parameters to match specified statistics. In their case, however, their constraint was a log-normal PDF of the pseudo-dissipation, which is not an analytically-derived property of isotropic turbulence. In fact, dissipation and enstrophy tend to exhibit stretched-exponential distributions \citep{Meneveau1991,Donzis2008}. The present model contrasts that of Girimaji and Pope by closing the conditional averages through physically-motivated assumptions supplemented by analytic constraints rather than by ad-hoc tuning to create a specific form of the dissipation PDF. In fact, the only empirical data entering the model comes from the need to specific the Gaussian coefficient $\delta$ using the skewness coefficient.

\section{Numerical Methods\label{sec_numerical}}

\subsection{Stochastic Differential Equation Solver\label{sec_SDE_numerical}}

The three models introduced in the previous sections (RFD, EGF and RDGF) can be advanced numerically as a system of stochastic ODEs. A second-order predictor-corrector method is used for time advancement. Time steps of size $dt/\tau_\eta = 0.04$, $0.02$, and $0.01$ are compared to verify discretization convergence. Ensembles of $2^{16}$ trajectories are advanced for $1000 \tau_\eta$ to achieve convergence of desired statistical quantities (up to fourth-order moments). Without loss of generality, $\tau_\eta = 1$ was used for all runs. The Fortran simulations are performed in serial and run on a desktop machine, taking a few hours to complete.

\subsection{Direct Numerical Simulation Database}

The Johns Hopkins Turbulence Databases (JHTDB) isotropic dataset \citep{Li2008,Perlman2007} provided the DNS statistics used for most of the comparisons in this paper. The dataset contains the simulation results from a $\Rey_\lambda = 430$ simulation of Navier-Stokes with forcing at the two lowest wavenumbers. The pseudo-spectral simulation provided a $1024^3$ resolution on a $(2\pi)^3$ cubic domain. Time advancement was accomplished via the $2^{nd}$-order Adams-Bashforth scheme and de-aliasing was done with $2\sqrt{2}/3$ truncation and random phase shift. In a few cases, the comparisons are supplemented with another DNS at $\Rey_\lambda = 160$ using the same simulation code. Important parameters for the simulations are given in Table \ref{tab_DNS}. It is worth noting that RFD model with $\tfrac{\tau_K}{T} = 0.1$ has been equated with $\Rey_\lambda = 150$ \citep{Chevillard2008}. Reaching $\Rey_\lambda = 430$ requires $\tfrac{\tau_K}{T} \approx 0.035$, which is outside the range for which RFD model produces results with reasonable accuracy. Therefore, in this paper, we use $\tfrac{\tau_K}{T} = 0.1$ for the RFD simulations, the value at which the RFD model seems to perform the best.

\begin{table}
%\def~{\hphantom{0}}
%\begin{ruledtabular}
\begin{center}
\begin{tabular}{c c c c c c c c}
%\hline
N & $\Rey_\lambda$ & $\epsilon$ & $\nu$ & $\eta$ & $\tau_\eta$ & $\Delta t$ & $k_{max}\eta$ \\
\hline
$256^3$ & $160$ & $0.112$ & $1.2$e-$03$ & $1.11$e-$02$ & $0.104$ & $5$e-$04$ & $1.34$ \\
$1024^3$ & $430$ & $0.093$ & $1.85$e-$04$ & $2.87$e-$03$ & $0.045$ & $2$e-$04$ & $1.39$ \\
%\hline
\end{tabular}
\caption{Numerical details for simulations used in this paper.}
%\end{ruledtabular}
\label{tab_DNS}
\end{center}

\end{table}

\section{Results \label{sec_results}}

\subsection{Longitudinal and Transverse Components}

\begin{figure}
\begin{center}
\includegraphics[width=0.49\linewidth]{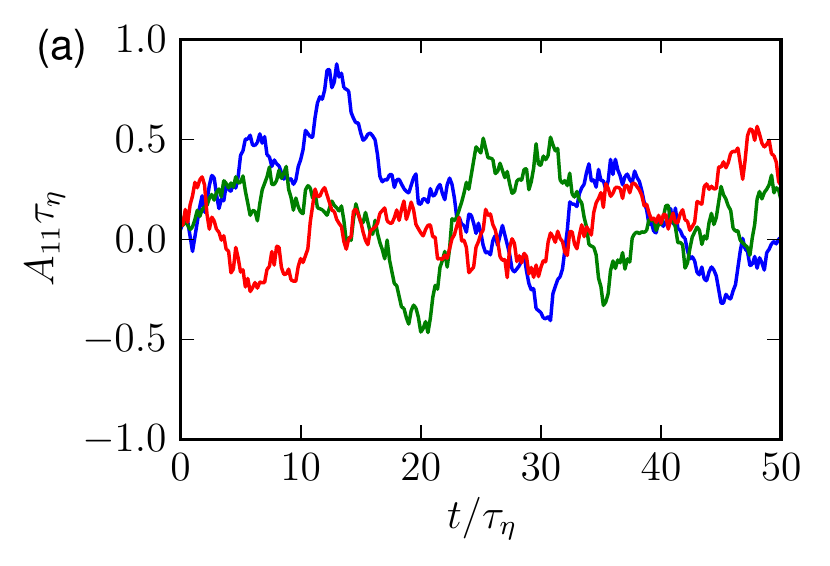}
\includegraphics[width=0.49\linewidth]{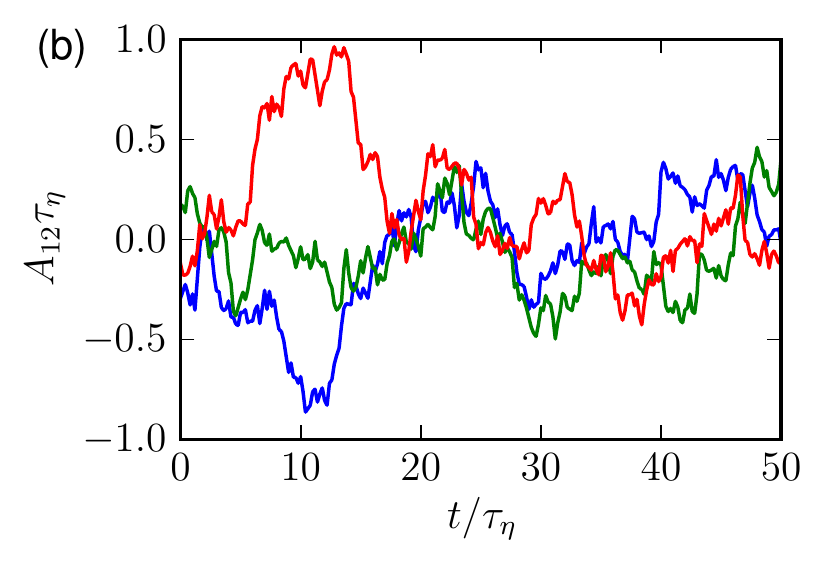}
\caption{Sample trajectories of (a) longitudinal and (b) transverse velocity gradient components from the RDGF mapping closure.}
\label{fig_samples}
\end{center}
\end{figure}

Figure \ref{fig_samples} illustrates the output of the RDGF mapping closure by plotting sample trajectories of longitudinal and transverse velocity components over an interval of $20 \tau_\eta$. Because of the stochastic forcing, the paths appear rough, even at the scale of the Kolmogorov timescale. Nonetheless, such stochastic models can be useful when their statistical behavior provides a good model for Lagrangian velocity gradient statistics in isotropic turbulence.

\begin{figure}
\begin{center}
\includegraphics[width=0.99\linewidth]{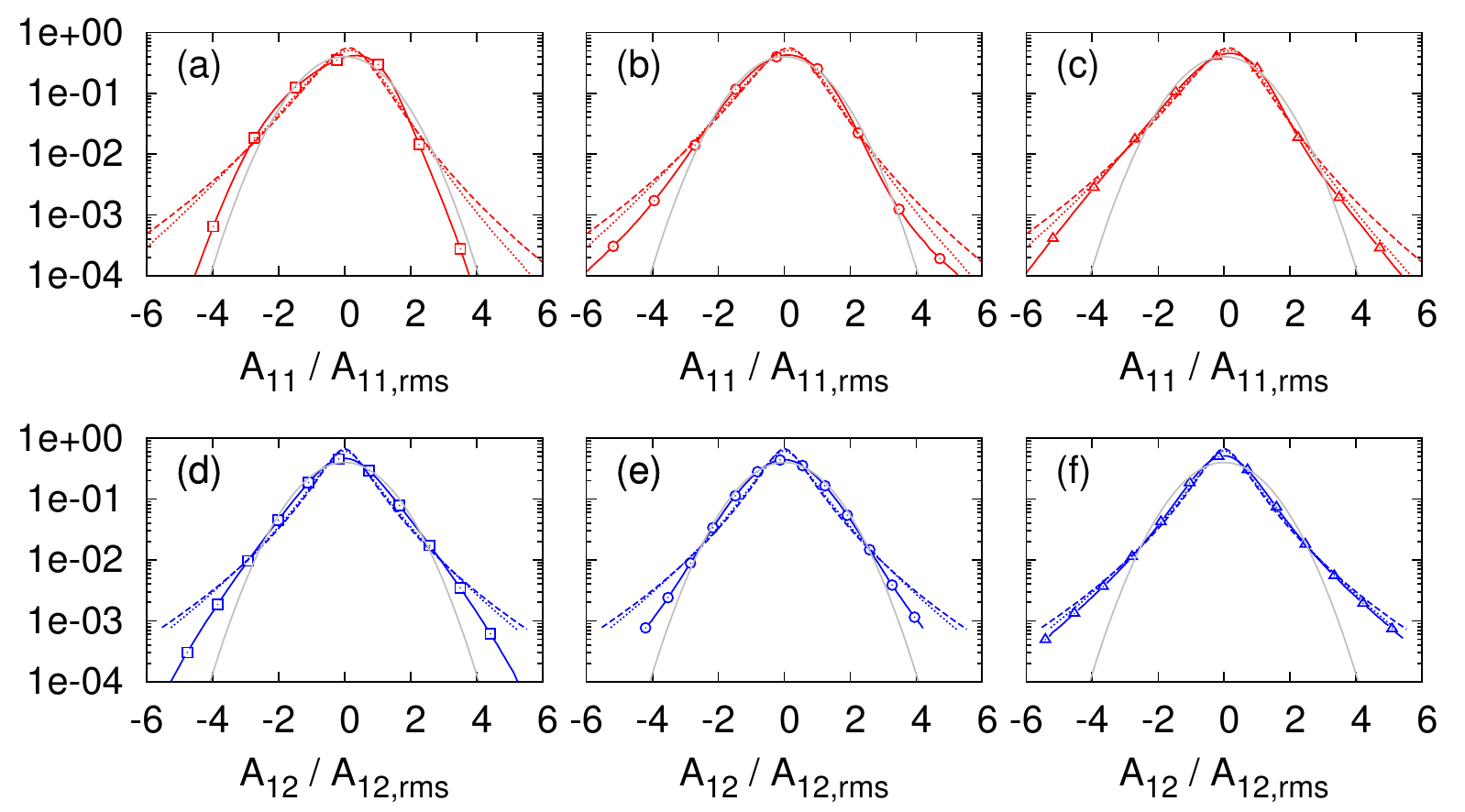}
\caption{Single component PDFs for longitudinal (a-c) and transverse (d-f) velocity components. Three models are compared: (a,d) RFD, (b,e) EGF, (c,f) RDGF mapping closure. Solid gray line denotes Gaussian, dashed line shows DNS results at $\Rey_\lambda = 430$, dotted line shows DNS at $\Rey_\lambda = 160$, and solid line with markers shows the model result.}
\label{fig_A11_A12}
\end{center}
\end{figure}

The probability density functions for the longitudinal velocity derivative, $A_{11}$, and transverse velocity derivative, $A_{12}$, are shown in figure \ref{fig_A11_A12}. The RFD, EGF, and RDGF closures are compared with DNS results at the two different Reynolds numbers. The negative skewness expected for $A_{11}$ and the symmetry expected for $A_{12}$ are reflected by all three models. The RFD results appear much too close to Gaussian when compared with DNS results. The longitudinal velocity gradient distributions (top row of figure) from the EGF and RDGF models are better than that of RFD in terms of deviation from Gaussian behavior. For the transverse component, the RFD and EGF results appear similar, being between Gaussian and the DNS results. The RDGF mapping closure provides a much better match for the $A_{12}$ PDF. The trends suggest that the RDGF model may provide an even better fit for DNS data at slightly lower $\Rey_\lambda$, but we refrain from any iterative matching with any particular precise value of  $\Rey_\lambda$ as we are mostly interested in overall trends.

As a compact comparison, Table \ref{tab_skewness_kurtosis} records the skewness and flatness factors of the above PDFs. All three models significantly under-predict the magnitude of the negative skewness for $A_{11}$, though the RFD and RDGF mapping closures are much closer than the EGF closure. The flatness factors for the longitudinal and transverse components help quantify the tendency of the model to reproduce the fattened tails of the PDFs in figure \ref{fig_A11_A12}. For the longitudinal component, the EGF model appears to give the closest match, while RDGF is slightly closer for the transverse component. In each case, the flatness factors are too low, as was probably already evident in the above figures. It appears that the {\it trend} in the RFD and RDGF mapping closures that the longitudinal component has lower flatness than the transverse component better reflects the DNS trend. Indeed, as was discussed above, these results for RDGF could be seen as somewhat more appropriate for matching the DNS results at even lower Reynolds number.

\begin{table}
\begin{center}
\begin{tabular}{c c c c c}
~ & $ \frac{\langle A_{11}^3 \rangle}{\langle A_{11}^2 \rangle^{3/2}} $ & $ \frac{\langle A_{12}^3 \rangle}{\langle A_{12}^2 \rangle^{3/2}}$ & $ \frac{\langle A_{11}^4 \rangle}{\langle A_{11}^2 \rangle^{2}} $  & $ \frac{\langle A_{12}^4 \rangle}{\langle A_{12}^2 \rangle^2} $ \\
\hline
RFD & -0.44 & 0.0 & 3.2 & 4.3 \\
EGF  & -0.31 & 0.0 & 6.5 & 6.3 \\
RDGF & -0.45 & 0.0 & 4.7 & 6.8 \\
$\Rey_\lambda = 160$  & -0.52 & 0.0 & 5.9 & 9.4 \\
$\Rey_\lambda = 430$  & -0.60 & 0.0 & 8.5 & 13.2 \\
\end{tabular}
\caption{Skewness and kurtosis values for longitudinal and transverse velocity gradient components from each model compared with DNS.}
\label{tab_skewness_kurtosis}
\end{center}
\end{table}

\subsection{Isotropic Relations}

Table \ref{tab_isotropic_relations} compares the extent to which each of the models is able to reproduce important isotropy relations. Each ratio is equal to unity for isotropic turbulence. The first ratio, $\frac{\langle S_{ij} S_{ij}\rangle}{\langle\Omega_{ij}\Omega_{ij}\rangle}$, represents the ratio of strain-rate magnitude to vorticity magnitude produced by the model and is equal to unity since by construction (adjustment of forcing parameters), $\langle Q \rangle = 0$. The second ratio, $\frac{-\tfrac{1}{3}\langle S_{ij} S_{jk} S_{ki} \rangle}{\tfrac{1}{4}\langle \omega_i S_{ij} \omega_j \rangle}$, represents the balance between strain production and vorticity production and is equal to unity if $\langle R \rangle = 0$, also expected due to the adjustment of forcing parameters. The identities  are all satisfied within numerical error showing that the numerical tuning of the three parameters ($D_s$,$D_a$, and $\tau$) is very accurate.  This represents a significant advantage of the RDGF mapping closure, seeing that the earlier RFD model slightly over-emphasizes strain-dominant and strain-production-dominant regions while the EGF model significantly over-emphasizes rotation-dominant and rotation-production-dominant regions. All three models satisfy the relation between dissipation and the longitudinal velocity derivative variance.

\begin{table}
\begin{center}
\begin{tabular}{c c c c c}
~ & $\frac{\langle S_{ij} S_{ij}\rangle}{\langle\Omega_{ij}\Omega_{ij}\rangle} $ & $\frac{-\tfrac{1}{3}\langle S_{ij} S_{jk} S_{ki} \rangle}{\tfrac{1}{4}\langle \omega_i S_{ij} \omega_j \rangle}$ & $\frac{15 \langle A_{11}^2 \rangle}{2 \langle S_{ij} S_{ij}\rangle}$ & $\frac{-\tfrac{35}{2} \langle A_{11}^3 \rangle}{\langle \omega_i S_{ij} \omega_j \rangle}$ \\
\hline
RFD & 1.143 & 1.76 & 1.00 & 1.76 \\
EGF  & 0.486 & 0.52 & 1.00 & 0.46 \\
RDGF & 1.000 & 1.00 & 1.00 & 1.00
\end{tabular}
\caption{Results for competing models in terms of reproducing known isotropic relations.}
\label{tab_isotropic_relations}
\end{center}
\end{table}

\subsection{Enstrophy and Dissipation}

\begin{figure}
\begin{center}
\includegraphics[width=0.99\linewidth]{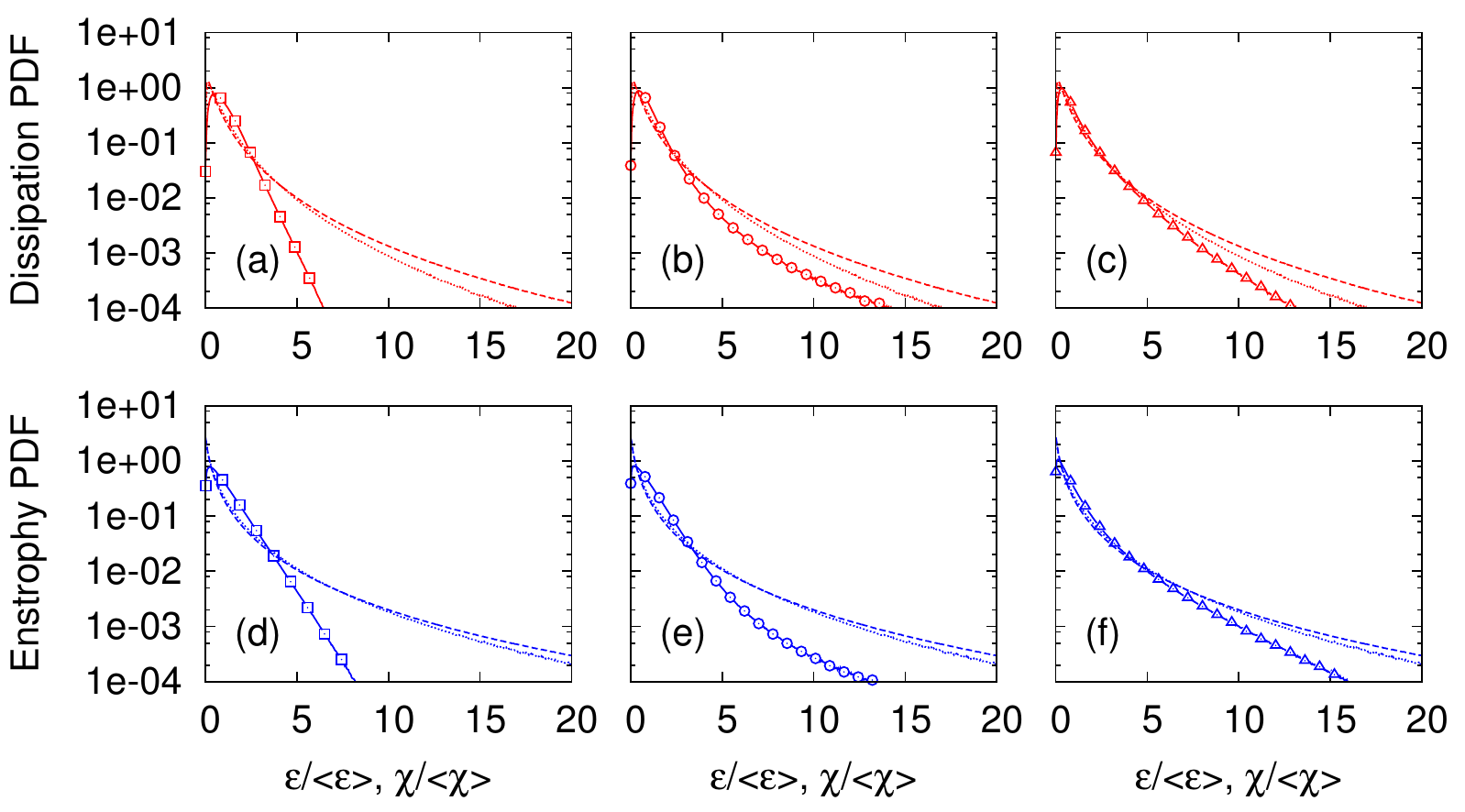}
\caption{PDFs of dissipation (a-c) and enstrophy (d-f) normalized by their mean values for RFD (a,d), EGF (b,e), RDGF (c,f). Solid lines with symbols indicate model results, and DNS results are shown with dashed ($\Rey_\lambda = 430$) and dotted ($\Rey_\lambda = 160$) lines.}
\label{fig_dissipation_enstrophy_pdf}
\end{center}
\end{figure}

The probability density distributions (PDFs) of enstrophy and dissipation  in isotropic turbulence \citep{Meneveau1991,Bershadskii1993,Donzis2008} provide  another useful test for comparing Lagrangian velocity gradient models. Figure \ref{fig_dissipation_enstrophy_pdf} compares the dissipation (top) and enstrophy (bottom) PDFs of the RFD (left), EGF (middle), and RDGF (right) models with the DNS results at two $\Rey_\lambda$ values. The RFD model appears to produce exponential tails (straight lines on the log-linear plot) rather than stretched exponential. The EGF model is much improved for the dissipation and enstrophy PDF, appearing somewhat closer to the characteristic stretched exponential shape. The RDGF model provides the best agreement with both dissipation and enstrophy distributions, displaying the stretched-exponential shape for both. It should be kept in mind that the EGF and RDGF do not have explicit Reynolds number dependence. Again, as a qualitative observation, the RDGF model gives results that may appear even more realistic for lower $\Rey_\lambda$.

\subsection{Vorticity and Strain-Rate}

One of the well-known features of velocity gradient statistics in turbulent flows is the non-trivial alignment of the vorticity vector with respect to the three eigenvectors of the strain-rate tensor \citep{Ashurst1987}. The vorticity tends to align more closely with the strain-rate eigenvector associated with the intermediate eigenvalue. Meanwhile, the vorticity tends to be more perpendicular with respect to the strain-rate eigenvector of the smallest eigenvalue. The alignment distribution between the vorticity and the eigenvector of the largest strain-rate eigenvalue tends to be fairly uniform in comparison.

\begin{figure}
\begin{center}
\includegraphics[width=0.99\linewidth]{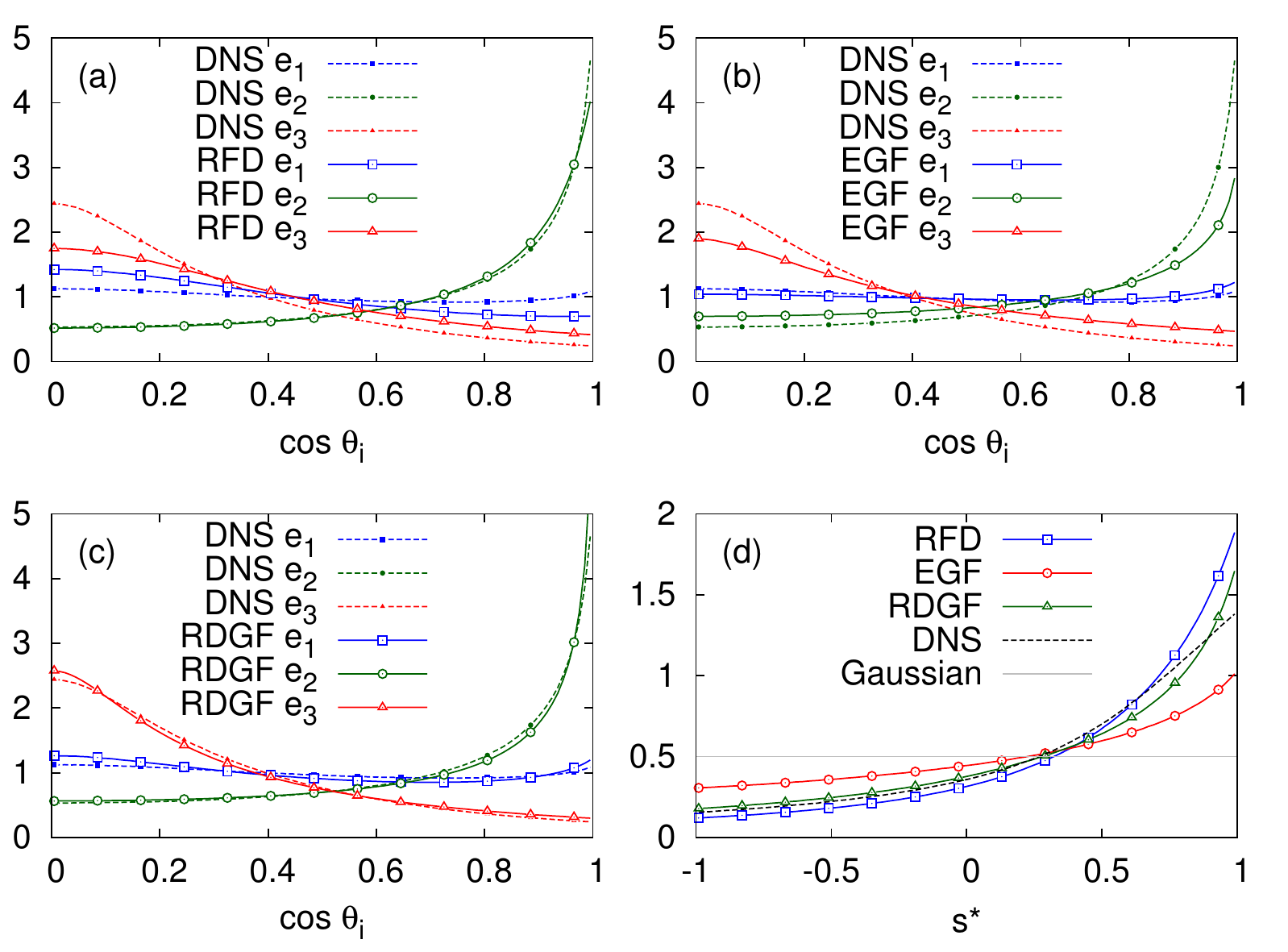}
\caption{Probability distribution functions for the cosine of the angle between vorticity and the strain-rate eigenvectors: (a) RFD, (b) EGF, (c) RDGF. (d) Probability density functions for $s^*$, as defined in  \eqref{eq_sstar}, for the three models compared with DNS results and Gaussian field statistics.}
\label{fig_align}
\end{center}
\end{figure}

Figure \ref{fig_align}a-c shows the PDFs for the cosines of the angles between vorticity and strain-rate eigenvectors. The DNS results at $\Rey_\lambda = 430$ are used for comparison here; these statistics show virtually no dependence on $\Rey_\lambda$. All three models mimic the well-known trend outlined above. The RFD model slightly underpredicts the anti-alignment of vorticity with the smallest strain-rate eigenvalue, while displaying a slight preference toward anti-alignment for the largest eigenvalue. The EGF consistently under-predicts the alignment biases seen in the DNS results. It appears that the RDGF model obtains the best agreement overall.

\citet{Lund1994} introduced the measure $-1 \leq s^* \leq 1$ using the eigenvalues of the strain-rate tensor,
\begin{equation}
s^* = - \frac{3 \sqrt{6} \Lambda_1 \Lambda_2 \Lambda_3}{\left( \Lambda_1^2 + \Lambda_2^2 + \Lambda_3^2 \right)^{3/2}},
\label{eq_sstar}
\end{equation}
which compares the relative magnitudes of each of the three strain-rate eigenvalues taking into account that they must add up to zero.
%The PDF of $s^*$ is uniform for Gaussian fields. The limit $s^* = -1$ represents equal contraction along two eigen-directions with incompressibility-conserving expansion along the third, i.e. $\Lambda_1:\Lambda_2:\Lambda_3 = 2:-1:-1$, distortion toward a cigar-shaped fluid volume. The other limit, $s^* = 1$, occurs for equal expansion along two eigen-directions with incompressibility-conserving contraction along the third, i.e. $\Lambda_1:\Lambda_2:\Lambda_3 = 1:1:-2$, distortion toward a disk-shaped fluid volume. The middle, $s^* = 0$, represents equal contraction and expansion along two of the eigen-directions with zero component along the third, i.e. $\Lambda_1:\Lambda_2:\Lambda_3 = 1:0:-1$.
Figure \ref{fig_align}(d) reports the PDFs for the three models considered here, shown in comparison to DNS results ($\Rey_\lambda = 430$). It is well-known that turbulent velocity gradients are biased toward $s^* > 0$, i.e. more distortion toward disk-like fluid elements \citep{Lund1994,Meneveau2011}. All three models reflect this trend. The RFD model over-predicts the bias toward positive $s^*$, while the EGF model under-predicts it. The RDGF model appears to produce results in closest comparison with DNS.

Table \ref{tab_strain_rate_vorticity} compares ensemble averages for some of these vorticity and strain-rate statistics, helping quantify the above discussion. Additionally available from this table is the ratio of average strain-rate eigenvalues, for which the RDGF models also provides good predictions.

\begin{table}
\begin{center}
\begin{tabular}{c c c c c c c}
~ & $ \langle s^* \rangle $ & $ \langle \Lambda_1 \rangle \tau_\eta$  & $\frac{\langle \Lambda_2 \rangle}{\langle \Lambda_1 \rangle}$  & $\langle \cos(\theta_1) \rangle$ & $\langle \cos(\theta_2) \rangle$ & $\langle \cos(\theta_3) \rangle$ \\
\hline
RFD & 0.441 & 0.400 & 0.270 & 0.428 & 0.663 & 0.374 \\
EGF  & 0.190 & 0.421 & 0.123 & 0.500 & 0.597 & 0.377 \\
RDGF & 0.347 & 0.392 & 0.224 & 0.473 & 0.656 & 0.317 \\
DNS  & 0.371 & 0.366 & 0.231 & 0.484 & 0.659 & 0.311
\end{tabular}
\caption{Various mean values for strain-rate and vorticity measures.}
\label{tab_strain_rate_vorticity}
\end{center}
\end{table}

\subsection{Dynamics in the Q-R Plane}

\begin{figure}
\begin{center}
\includegraphics[width=0.99\linewidth]{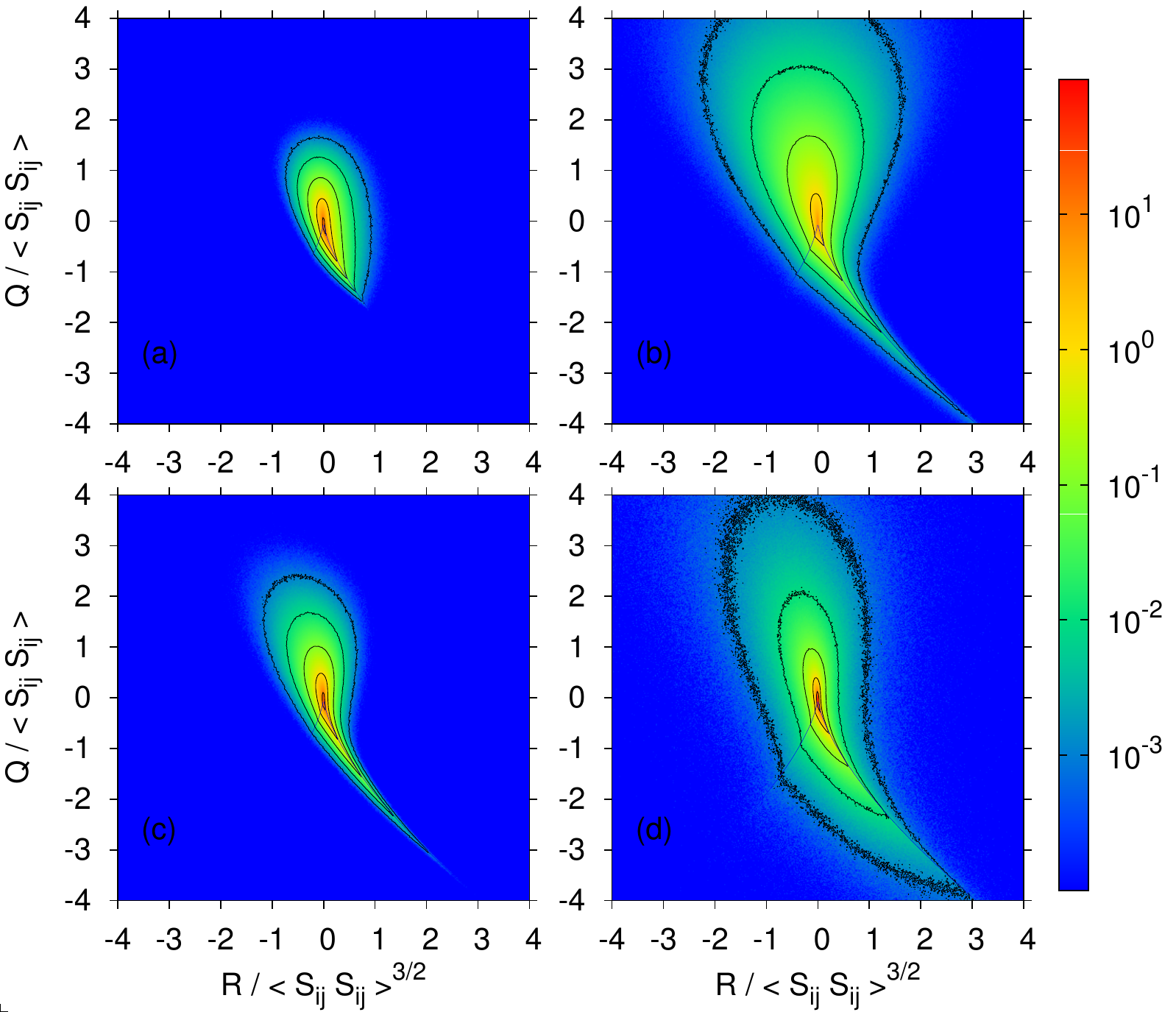}
\caption{Logarithmically scaled joint-probability density function for the invariants $Q$ and $R$ as given by (a) RFD, (b) EGF, (c) RDGF, and (d) DNS.}
\label{fig_QRpdf}
\end{center}
\end{figure}

Another salient feature of turbulent velocity gradient statistics is the teardrop shaped contours of the joint-probability density function for the $Q$ and $R$ invariants. Figure \ref{fig_QRpdf} compares such joint PDFs from the three models with DNS results ($\Rey_\lambda = 430$). Each model reproduces to some extent the features in the DNS results, most notably the teardrop shape.

The RFD results are too compact, lacking sufficient excursions far from the mean, as also seen previously for the single component PDFs in figure \ref{fig_A11_A12}. One also observes a less prominent high-probability filament descending down the positive $R$ branch of the Viellefosse line. The EGF model results are more accurate in their depiction of the high probability region along the Viellefosse line but a less realistic aspect of the EGF results is the exaggerated higher-probability in the positive $Q$ region compared to the negative $Q$ region. This feature is evidently responsible for the EGF model's departure from $\langle Q \rangle = 0$ (the EGF also does not reproduce $\langle R \rangle = 0$). The results from the RDGF mapping closure share some of these strengths and weaknesses. For the RDGF, the low probability contours remain too compact, though less so than in the case of the RFD model. The shape of the high-probability regions closely mirrors those for the DNS. Additionally, there is some promising spread for the low-probability contours into the high positive $Q$ regions. However, overall, the details of the low-probability contours (the tails of the joint distribution) still represents a challenge for all three models.

Neglecting the stochastic forcing for the moment, the dynamical equations for $Q$ and $R$ are \citep{Chevillard2008},
\begin{equation}
\frac{dQ}{dt} = -3R + A_{ij} P_{ji}^{(d)} - \nu A_{ij} \nabla^2 A_{ji}, \hspace{0.05\linewidth} \frac{dR}{dt} = \frac{2}{3} Q^2 + A_{ij} A_{jk} P_{ki}^{(d)} - \nu A_{ij} A_{jk} \nabla^2 A_{ki}.
\end{equation}
The dynamics in probability space can be recovered thus from conditional averaging,
\begin{equation}
\left\langle \left. \frac{dQ}{dt} \right| Q, R \right\rangle = - 3 R + \left\langle \left. A_{ij} P_{ji}^{(d)}  \right| Q, R \right\rangle - \nu \left\langle \left. A_{ij} \nabla^2 A_{ji}  \right| Q, R \right\rangle,
\end{equation}
\begin{equation}
\left\langle \left. \frac{dR}{dt} \right| Q, R \right\rangle = \frac{2}{3} Q^2 + \left\langle \left. A_{ij} A_{jk} P_{ki}^{(d)}  \right| Q, R \right\rangle - \nu \left\langle \left. A_{ij} A_{jk} \nabla^2 A_{ki}  \right| Q, R \right\rangle.
\end{equation}
These equations represent average velocities in the $QR$ probability space which, when multiplied with the local probability density,  represent fluxes in probability space. They are evaluated based on DNS as well as from the three models. In order to compare them under similar conditions,  averages are evaluated as an {\it a priori} test, by evaluating the model results from an ensemble of DNS trajectories. In practice, we found that the most significant effect of this approach (as opposed to sample the statistics along model evaluations) was to increase the domain in $QR$ space where the average velocities could be obtained.

\begin{figure}
\begin{center}
\includegraphics[width=0.49\linewidth]{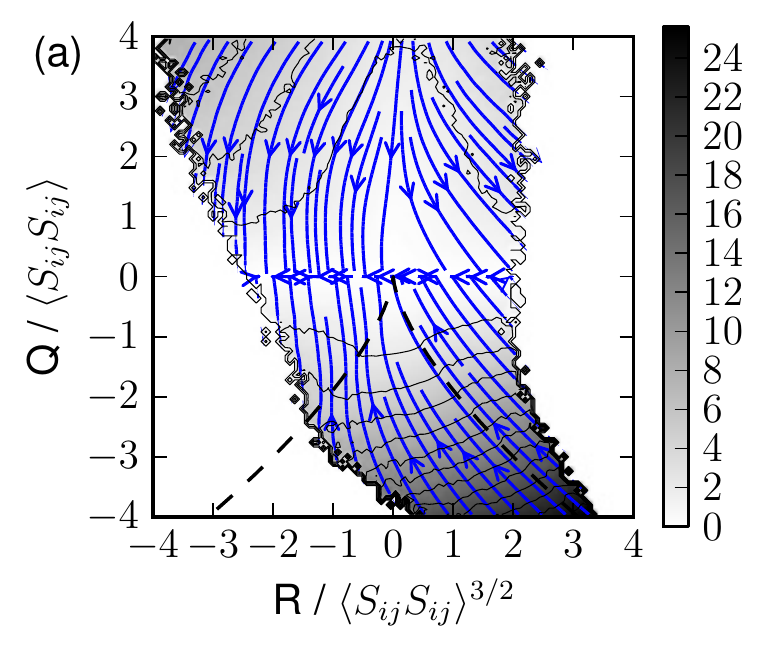}
\includegraphics[width=0.49\linewidth]{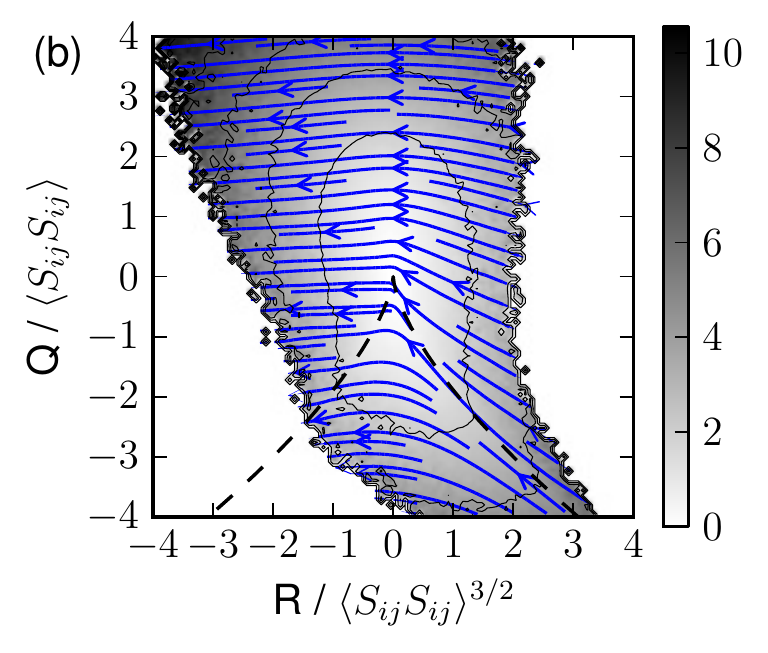}\\
\includegraphics[width=0.49\linewidth]{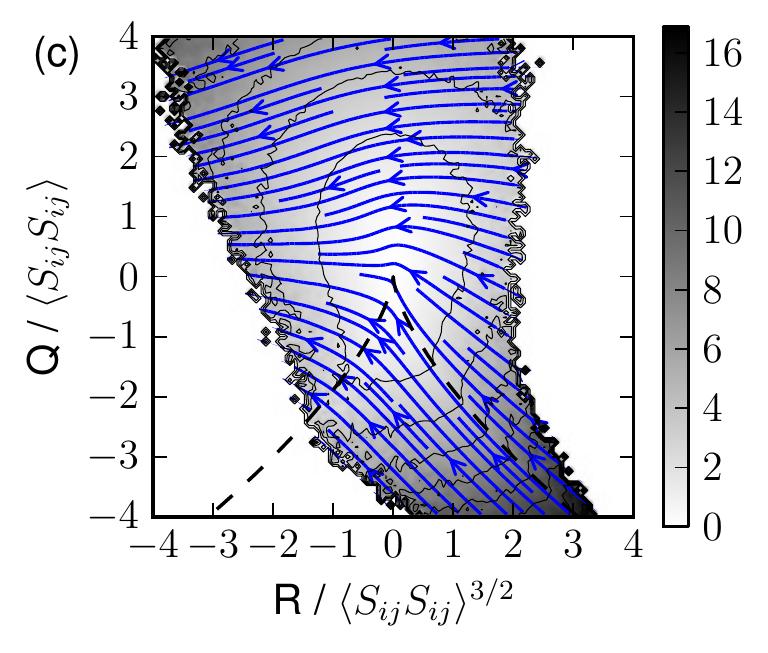}
\includegraphics[width=0.49\linewidth]{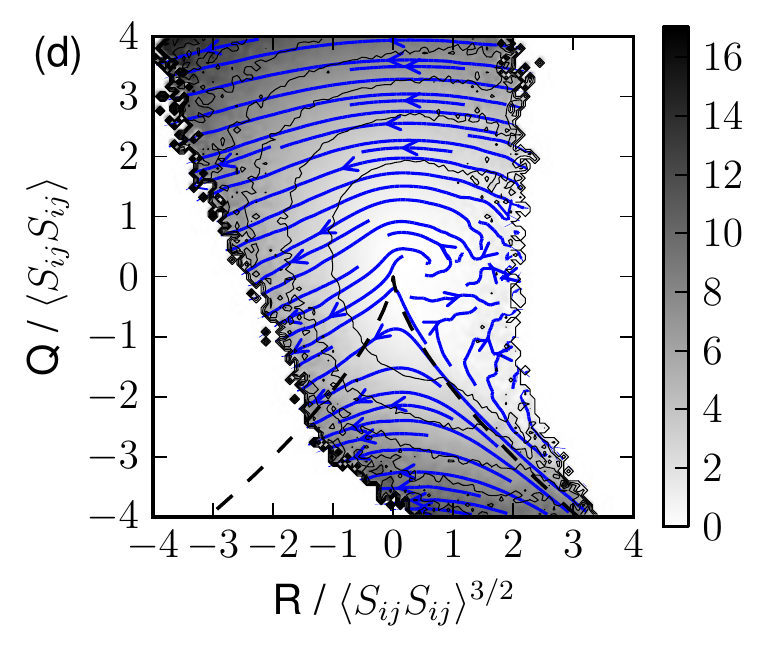}\\
\caption{Thick lines with arrows represent ``streamlines'' in the $QR$-plane due to the deviatoric part of the pressure Hessian. Thin lines represent contours for the velocity magnitude in the $QR$-plane. Results are as given by (a) RFD, (b) EGF, (c) RDGF mapping closure, and (d) DNS.}
\label{fig_QRvel_PHd}
\end{center}
\end{figure}

Figure \ref{fig_QRvel_PHd} shows the $QR$-space velocities attributed to the pressure Hessian term for the three models compared with DNS results ($\Rey_\lambda = 430$). The primary action of the RFD pressure Hessian is to oppose the restricted Euler motion along the Vieillefosse tail. In fact, the magnitude of the pressure Hessian opposing the restricted Euler singularity along the Vieillefosse tail is too strong in comparison with the DNS data. As previously noted \citep{Chevillard2008}, the RFD pressure Hessian lacks the right-to-left motion seen in the DNS and the other two models. This elucidates the shortcoming of the upstream isotropic assumption for the pressure Hessian tensor. In fact, it is a significant contribution of the Gaussian form of the pressure Hessian that it adds this right-to-left tendency due to the deviatoric component of the tensor.

The EGF pressure Hessian tends to oppose the singularity with smaller magnitude than the DNS results indicate, while the RDGF opposes with slightly larger magnitude than DNS. While the right-to-left motion is captured by the EGF and RDGF closures, a few more subtle features of the DNS results are not. First, the relatively ambient region of positive $R$ near $Q = 0$ has an unphysically active right-to-left motion in the EGF and RDGF closures. Secondly, the DNS results indicate opposition to restricted Euler along the left side of the Vieillefosse line, which is not replicated by the EGF or RDGF closures. Other subtle differences and similarities may be noted, but the above discussion summarizes the most important trends noticeable.

\begin{figure}
\begin{center}
\includegraphics[width=0.49\linewidth]{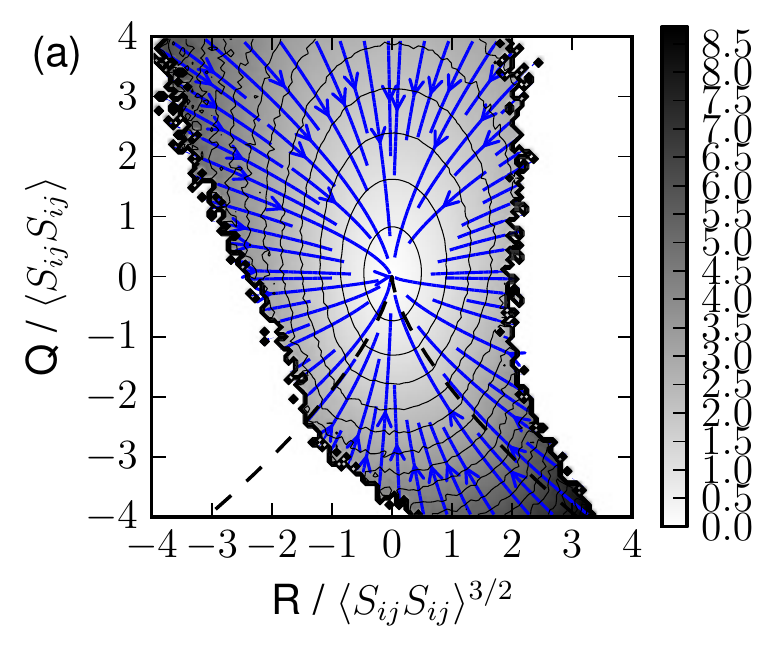}
\includegraphics[width=0.49\linewidth]{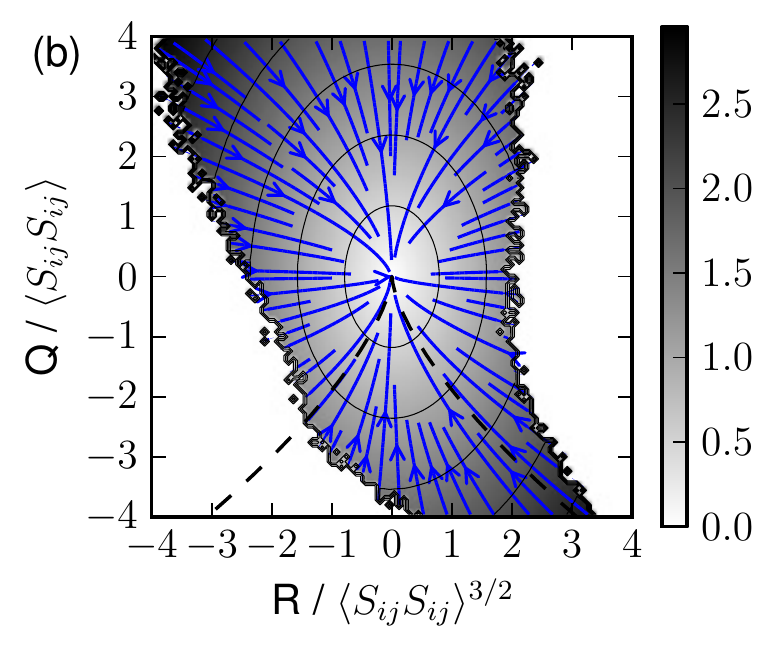}\\
\includegraphics[width=0.49\linewidth]{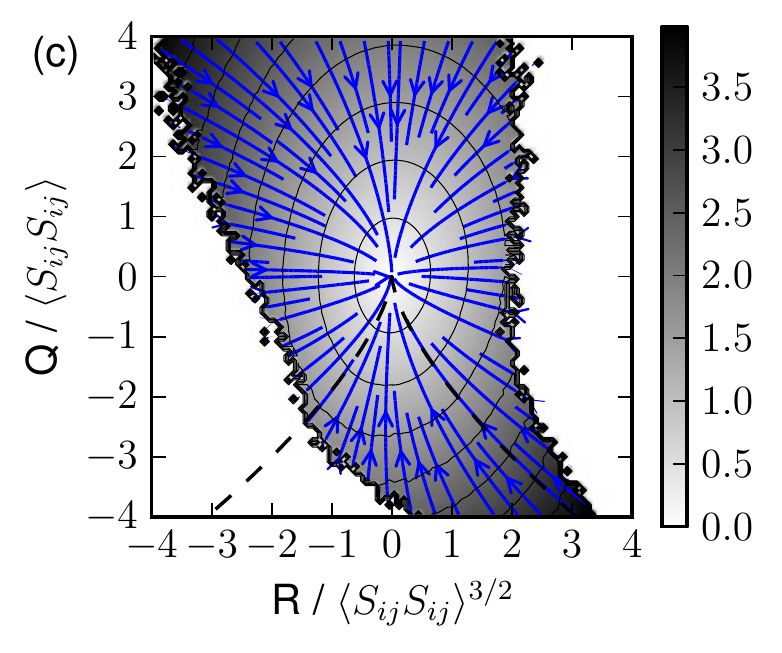}
\includegraphics[width=0.49\linewidth]{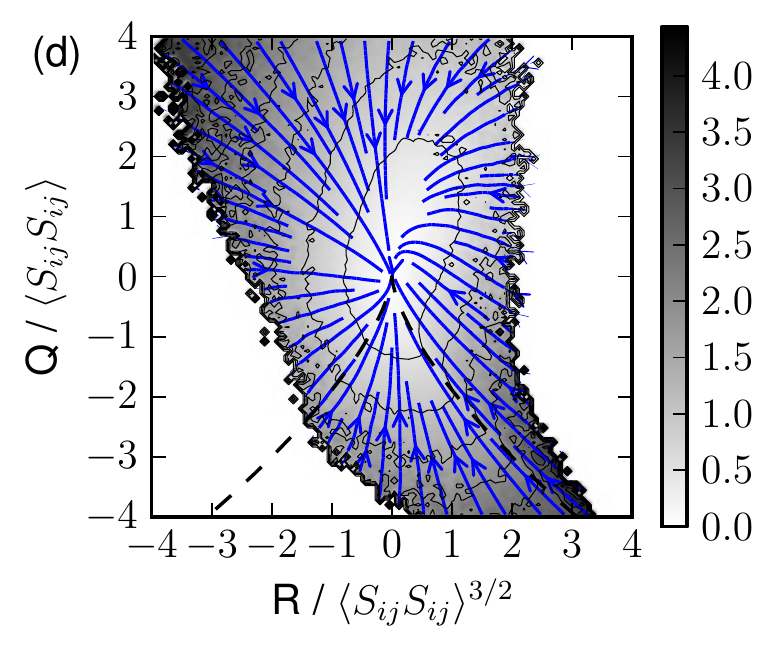}\\
\caption{Thick lines with arrows represent streamlines in the $QR$-plane due to the viscous Laplacian. Thin lines represent contours for the velocity magnitude in the $QR$-plane, non-dimensionalized by powers of $\left\langle S_{ij} S_{ij} \right\rangle$. Results are as given by (a) RFD, (b) EGF, (c) RDGF, and (d) DNS. }
\label{fig_QRvel_VL}
\end{center}
\end{figure}

The velocities in $QR$-space from the viscous Laplacian are shown in figure \ref{fig_QRvel_VL} for each of the models compared with DNS. All the models produce the same structure: the viscous Laplacian damps the velocity gradient, thus trajectories are pushed toward the origin in $QR$-space. Note that the DNS results show some slight deviation from pure damping structure. For example, near $Q = 0$ for $R > 0$, there is an upward trend in the streamlines instead of proceeding straight toward the origin. Each of the models fail to capture this effect. Thus, updating the upstream conditions of the conditional viscous Hessian produces minimal changes in the behavior of the closure. It appears that the upstream isotropic assumption of RFD model for the viscous term produces relatively more accurate results than was the case for the pressure Hessian.

In terms of magnitude, the RFD model is too strong. The EGF model produces good agreement with DNS in magnitude for the $Q < 0$, $R > 0$ region near the Vieillefosse tail, while it is too weak in the $Q > 0$, $R < 0$ region. The RDG model has magnitudes in good agreement with DNS for $Q > 0$, $R < 0$ but is too strong in the $Q < 0$, $R > 0$ along the Viellefosse tail. 

The above $QR$-space  analysis shows advantages of the EGF and RDGF closures over the RFD closure. Of particular importance is that the RFD pressure Hessian does not have a strong tendency to decrease $R$. The structure of the deviatoric pressure Hessian from the Gaussian fields provides this effect. Furthermore, the RFD model's over-prediction of magnitude for both of the unclosed terms results in the overly compact joint-PDF contours seen in figure \ref{fig_QRpdf}.

\subsection{Correlation Coefficients}

It is interesting to compare the {\it a priori} success of each model in terms of correlation coefficients for the deviatoric part of the pressure Hessian and the viscous Laplacian. For the deviatoric part of the pressure Hessian, the correlation coefficient is defined as,
\begin{equation}
\rho_{\mathbf{P}^{(d)}} = \frac{\left\langle P_{ij}^{(d),\text{DNS}} P_{ij}^{(d),\text{model}} \right\rangle}{\sqrt{\left\langle P_{ij}^{(d),\text{DNS}} P_{ij}^{(d),\text{DNS}} \right\rangle \left\langle P_{ij}^{(d),\text{model}} P_{ij}^{(d),\text{model}} \right\rangle}}.
\end{equation}
A similar correlation coefficient is also defined for the viscous Laplacian. These are computed using $8^{th}$-order finite differencing from an ensemble of 10 million points in the DNS results.

Table \ref{tab_correlation_coeffs} shows the resulting correlation coefficients. Included also is the original Gaussian Fields (GF) closure of \citet{Wilczek2014}, which did not provide a statistically stationary solution but rather succumbs to the finite-time singularity similar to the restricted Euler. Overall, the viscous Laplacian models are more successful than the pressure Hessian models. The RFD model has the lowest {\it a-priori} correlation coefficients for both closures. The difference between the GF and EGF model in Table \ref{tab_correlation_coeffs} is minimal.
% CM: I am not sure about this, so I cut it out.., which highlights the fact that the correlation coefficient is more sensitive to the {\it structure} of the Gaussian model rather than the %{\it magnitude} of its coefficients. Indeed, the viability of a model depends on the ability to prevent singularities, which depends more on the magnitude than structure. Therefore, %while correlation coefficients can help to judge the accuracy of a model in conjunction with other measures, caution is to be used in interpreting the results.

The RDGF model actually shows slightly lower correlation for its pressure Hessian model, indicating that the effect of the recent deformation on the Gaussian structure is perhaps not as helpful as one might have hoped. Perhaps the real advantage of the recent deformation is that the magnitude is increased without abandoning the analytical coefficients (i.e. $\alpha = -\tfrac{2}{7}$, $\beta = \tfrac{2}{5}$). The effect is that the singularity is avoided without recourse to DNS-tuned coefficients. 

\begin{table}
\begin{center}
\begin{tabular}{c c c}
~ & $\rho_{\mathbf{P}^{(d)}} $ & $\rho_{\nabla^2\mathbf{A}}$  \\
\hline
RFD & 0.23 & 0.41  \\
GF   & 0.43 & 0.60 \\
EGF  & 0.43 & 0.60 \\
RDGF & 0.37 & 0.61
\end{tabular}
\caption{Correlation coefficients for three models with DNS results at $\Rey_\lambda = 430$.}
\label{tab_correlation_coeffs}
\end{center}
\end{table}

\subsection{Computational Cost}

It is useful to mention that these three models are not equal in terms of computational cost. The above results were computed using a Fortran 90 code executed on a single processor. A minimal code involving only time advancement of the velocity gradient tensor without any statistical calculations was timed for the three models. It was found that, per time step, the RFDF model requires about $1.5$ times longer than the EGF model, while the RDGF model takes about $2.5$ times longer. It is worth noting, however, that the RFD and RDGF models were found to run smoothly and accurately with a time step of about $dt = 0.04 \tau_\eta$, while the EGF model required a time step of $dt = 0.01 \tau_\eta$ to avoid singularity. Even with such a small time step, the stochastic system exhibited rare rogue trajectories that had an overwhelming effect on the flatness factors, preventing convergence in a reasonable amount of time (e.g. trajectories advanced for $1000 \tau_\eta$). We note that \citet{Wilczek2014} used an even smaller time step of $dt = 0.001 \tau_\eta$. Therefore, the computational cost advantage of EGF model is not realized. The RFD model does have a computation cost per time step approximately $40\%$ smaller than that of the RDGF model.

\section{Conclusions}

In this paper, a new closure, the Recent Deformation of Gaussian Fields (RDGF) mapping closure, for the pressure Hessian and viscous Laplacian along Lagrangian trajectories in turbulent flow is introduced. The new closure benefits from the insights of both the Recent Fluid Deformation (RFD) and Gaussian Field (GF) closures. The GF closure calculations are applied for the initial upstream conditions of the conditional pressure Hessian and viscous Laplacian, before performing a recent fluid deformation mapping to complete the closure. The coefficients for Gaussian fields can be used and three remaining free parameters related to forcing and time-scale are constrained so that the model reproduces known exact statistical relations. The stochastic forcing for this model is also generalized from that used for the previous models so that the magnitude of the symmetric and anti-symmetric forcings can be applied independently.

{\it A priori} evaluation of the models in terms of correlation coefficients and $QR$-space velocities reveals the shortcomings of RFD closure: the magnitudes of the unclosed terms are significantly over-estimated, and the role of the pressure Hessian in decreasing the $R$ invariant is absent. These shortcomings are much improved using the conditional pressure Hessian from Gaussian fields. On the other hand, the exponential non-linearity of the recent deformation tensor allows for more effective prevention of singularities. As a result, the RDGF model does not require DNS-tuned coefficients in order to prevent the singularity. In this way, the RDGF model has the robustness and analytical closedness of RFD model while providing a more realistic structure of the pressure Hessian from the GF closure.

A comparison of various single-time statistics suggests that the RDGF model can provide excellent results in comparison to the two previous models. However, by comparison with DNS at $\Rey_\lambda = 430$, the quantitative results reveal remaining shortcomings such as lack of increasing long tails and intermittency. The RDGF results seem more consistent with lower Reynolds number DNS results. This highlights one of the major limitations of the current model, that it does not include a robust way of changing the Reynolds number whereas velocity gradient statistics are known to depend strongly on Reynolds number. The RFD model does include a mechanism for increasing the Reynolds number, but only in a very limited range. In fact, RFD applied for $\Rey_\lambda \approx 430$ is already outside the range where it performs well. The RDGF mapping closure suffers these same drawbacks as RFD, even if the skewness factor is adjusted to reflect its (weak) dependence on Reynolds number.

In summary, this paper builds a new closure framework for the conditional pressure Hessian and viscous Laplacian which leverages insights of previous approaches. It provides, therefore, a promising direction for future investigations of velocity gradient statistics in isotropic turbulence. At sufficiently high Reynolds numbers, where approximate isotropy of small scales is a safe assumption, models for isotropic turbulence can be applicable for a more general class of turbulent flows, for which some applications may find efficient access to velocity gradient statistics useful.

\section*{Acknowledgments}

PJ was supported by a National Science Foundation Graduate Research Fellowship Program under Grant No.
DGE-1232825, and CM by a grant from the NSF CBET-1507469. 
PJ thanks Michael Wilczek for many insightful discussions during his time at Johns Hopkins.

\appendix

\section{Isotropic Tensorial Stochastic Forcing for Symmetric and Anti-Symmetric Components\label{app_forcing}}

In this appendix, the form of the stochastic forcing in  \eqref{eq_BG_LagraingianStochasticA} is established. As identified in the text, the forcing should have the form $dF_{ij} = b_{ijk\ell} dW_{k\ell}$, and can be thought of as a sum of symmetric and anti-symmetric forcing, $dF_{ij} = dF_{ij}^{(s)} + dF_{ij}^{(a)}$, where $dF_{ij}^{(s)} = \tfrac{1}{2} \left( dF_{ij} + dF_{ji} \right)$ and $dF_{ij}^{(a)} = \tfrac{1}{2} \left( dF_{ij} - dF_{ji} \right)$. Since $dW_{ij}$ represents a tensorial Wiener process, i.e. $\left\langle W_{ij} \right\rangle = 0$ and $\left\langle dW_{ij} dW_{k\ell} \right\rangle = \delta_{ik} \delta_{j\ell} dt$, then
\begin{equation}
\left\langle dF_{ij} dF_{k\ell} \right\rangle = b_{ijmn} b_{k\ell mn} dt = D_{ijk\ell} dt.
\end{equation}
Therefore, the forcing contributes a variance growth rate of
\begin{equation}
d \left\langle F_{ij} F_{ij} \right\rangle = \left\langle dF_{ij} dF_{ij} \right\rangle = D_{ijij} dt
\end{equation}
and furthermore, the symmetric and anti-symmetric variance growth rates are,
\begin{equation}
d \left\langle F_{ij}^{(s)} F_{ij}^{(s)} \right\rangle = \left\langle dF_{ij}^{(s)} dF_{ij}^{(s)} \right\rangle = \frac{1}{2} \left( D_{ijij} + D_{ijji} \right)  dt \equiv D_s dt.
\end{equation}
\begin{equation}
d \left\langle F_{ij}^{(a)} F_{ij}^{(a)} \right\rangle = \left\langle dF_{ij}^{(a)} dF_{ij}^{(a)} \right\rangle = \frac{1}{2} \left( D_{ijij} - D_{ijji} \right)  dt = D_a dt.
\end{equation}
Here, by definition, $D_s$ and $D_a$ represent the growth rate of the variance of symmetric and anti-symmetric parts of the forcing.

To model isotropic turbulence, the stochastic forcing should be statistically isotropic. The most general isotropic form for the diffusion tensor is
\begin{equation}
D_{ijk\ell} = d_1 \delta_{ij} \delta_{k\ell} + d_2 \delta_{ik} \delta_{j\ell} + d_3 \delta_{i\ell} \delta_{jk}.
\end{equation}
Requiring also that the forcing be trace-free (incompresibility), then
\begin{equation}
D_{iik\ell} = \left(3 d_1 + d_2 + d_3\right) \delta_{k\ell} = 0.
\end{equation}
Combining this constraint with the two definitions of $D_s$ and $D_a$ given above,
\begin{equation}
D_s = \frac{1}{2} \left( D_{ijij} + D_{ijji} \right) = 3 d_1 + 6 d_2 + 6 d_3,
\end{equation}
\begin{equation}
D_a = \frac{1}{2} \left( D_{ijij} - D_{ijji} \right) = 3 d_2 - 3 d_3,
\end{equation}
then the system of 3 equations and 3 unknowns can be solved for
\begin{equation}
D_{ijk\ell} = - \frac{D_s}{15} \delta_{ij} \delta_{k\ell} + \left( \frac{D_s}{10}  + \frac{D_a}{6} \right) \delta_{ik} \delta_{j\ell} + \left( \frac{D_s}{10} - \frac{D_a}{6} \right) \delta_{i\ell} \delta_{jk}.
\end{equation}
The choice of $D_s = D_a = 15$ reduces to the form of Chevillard and Meneveau \citep{Chevillard2008} used for the RFD model,
\begin{equation}
D_{ijk\ell} = - \delta_{ij} \delta_{k\ell} + 4 \delta_{ik} \delta_{j\ell} - \delta_{i\ell} \delta_{jk}.
\end{equation}

To implement this forcing, however, the tensor $b_{ijk\ell}$ is necessary, thus the equation $b_{ijmn}b_{k\ell mn} = D_{ijk\ell}$ must be solved. Using the general isotropic form
\begin{equation}
b_{ijk\ell} = b_1 \delta_{ij} \delta_{k\ell} + b_2 \delta_{ik} \delta_{j\ell} + b_3 \delta_{i\ell} \delta_{jk},
\end{equation}
the tensor contractions yield the following system of equations,
\begin{equation}
d_1 = 3 b_1^2 + 2 b_1 b_2 + 2 b_1 b_3 = - \frac{D_s}{15},
\end{equation}
\begin{equation}
d_2 = b_2^2 + b_3^2 = \frac{D_s}{10} + \frac{D_a}{6},
\end{equation}
\begin{equation}
d_3 = 2 b_2 b_3 = \frac{D_s}{10} - \frac{D_a}{6}.
\end{equation}
Solution to this system of equations yields,
\begin{equation}
b_{ijk\ell} = -\frac{1}{3} \sqrt{\frac{D_s}{5}} \delta_{ij} \delta_{k\ell} + \frac{1}{2} \left( \sqrt{\frac{D_s}{5}} + \sqrt{\frac{D_a}{3}} \right) \delta_{ik} \delta_{j\ell} + \frac{1}{2} \left( \sqrt{\frac{D_s}{5}} - \sqrt{\frac{D_a}{3}} \right) \delta_{i\ell} \delta_{jk},
\end{equation}
which reduces to the form of \citet{Chevillard2008} with the choice $D_s = D_a = 15$. Meanwhile, \citet{Wilczek2014} tuned $D_s = D_a$ such that the definition of $\tau_\eta$ was consistent between model and numerics. In the authors' current view, e.g. considering  \eqref{eq_BG_LagrangianStochasticS} and $\eqref{eq_BG_LagrangianStochasticW}$, there is no a-priori reason that the strain-rate and vorticity should be forced stocastically with the same amplitude, therefore, the present model considers $D_s$ and $D_a$ to be two independent tuning parameters.

\section{Analytical Calculation of $\gamma$ for the Gaussian Fields Representation of the Conditional Pressure Hessian\label{app_gamma}}

\begin{figure}
\begin{center}
\includegraphics{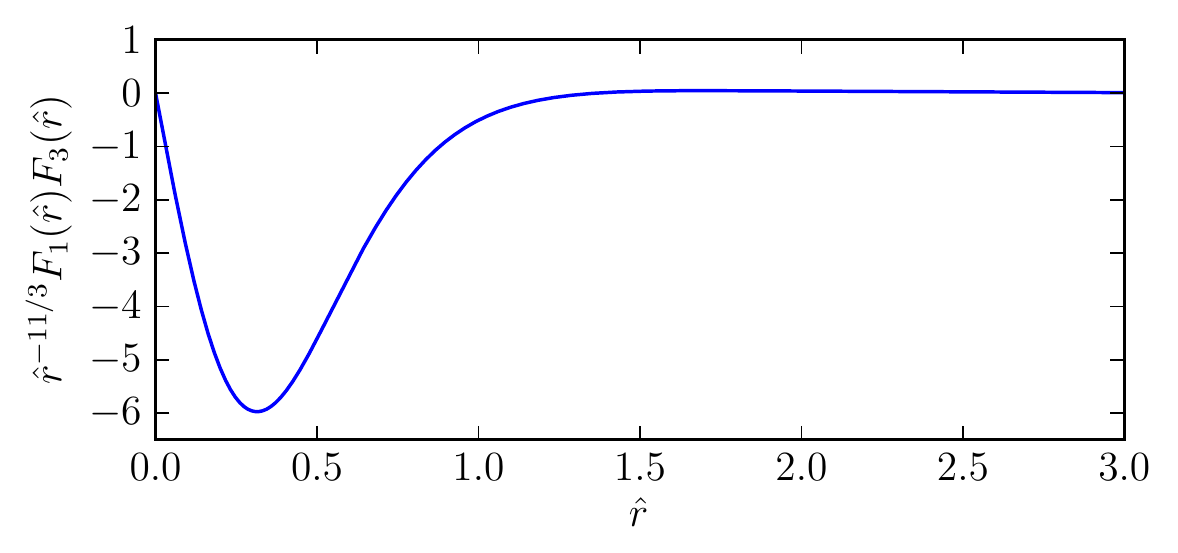}
\end{center}
\caption{Integrand in  \eqref{eq_app_gamma_integral} plotted in normalized variables $\hat{r} = \frac{r}{\gamma_2 \eta}$ with $\gamma_2 \approx 13$.}
\label{fig_app_gamma_integrand}
\end{figure}

A key component to both the Enhanced Gaussian closure and the recent-deformation of Gaussian fields mapping closure is the representation of a conditional pressure Hessian using  \eqref{eq_BG_GaussianPressureHessian}. While the coefficients $\alpha$ and $\beta$ were directly evaluated from the Gaussian fields closure, the last coefficient is determined by the details of the longitudinal correlation function,  \eqref{eq_BG_GaussianCoeffs}. Calculations are easier using the longitudinal structure function $D_{LL}(r) = 2\overline{u^2} (1 - f(r))$.
\begin{equation}
\gamma = \frac{6}{25} + \frac{16}{75 D_{LL}''(0)^2} \int\limits_{0}^{\infty} \frac{D_{LL}'(r)D_{LL}'''(r)}{r} dr,
\end{equation}
where $D_{LL}''(0) = \frac{2\epsilon}{15 \nu}$ according to the proper viscous range behavior. Using the approach of \citet{Batchelor1951}, the viscous and inertial range behavior of the structure function can be preserved using a blending function,
\begin{equation}
D_{LL}(r) = C_2 \epsilon^{2/3} r^{2/3} F\left(\frac{r}{\gamma_2\eta}\right).
\end{equation}
Here, we assume K41 scaling for the inertial range with Kolmogorov coefficient $C_2 \approx 2.0$ \citep{Pope2000}. The blending function of \citet{Batchelor1951} is
\begin{equation}
F\left(\frac{r}{\gamma_2\eta}\right) = \left[ 1 + \left( \frac{r}{\gamma_2 \eta} \right)^{-2} \right]^{-2/3},
\end{equation}
where $\gamma_2 = \left( 15 C_2 \right)^{3/4} \approx 13$ sets the cross-over point between viscous and inertial behavior, recovering the correct viscous range behavior. With the application of product rule differentiation, we can write
\begin{equation}
D_{LL}'(r) = C_2 \epsilon^{2/3} r^{-1/3} F_1\left( \frac{r}{\gamma_2 \eta} \right),
\end{equation}
\begin{equation}
D_{LL}'''(r) = C_2 \epsilon^{2/3} r^{-7/3} F_3\left( \frac{r}{\gamma_2 \eta} \right),
\end{equation}
and thus the integral simplifies under the change of variable $\hat{r} = \tfrac{r}{\gamma_2 \eta}$,
\begin{equation}
\gamma = \frac{6}{25} + \frac{12}{225} I,
\end{equation}
where
\begin{equation}
I = \int\limits_{0}^{\infty} \hat{r}^{-11/3} F_1(\hat{r}) F_3(\hat{r}) d\hat{r},
\label{eq_app_gamma_integral}
\end{equation}
with the derivative functions
\begin{equation}
F_1(\hat{r}) = \frac{2}{3} F(\hat{r}) + \hat{r} F'(\hat{r}),
\end{equation}
\begin{equation}
F_3(\hat{r}) = \frac{8}{27} F(\hat{r}) - \frac{2}{3} \hat{r} F'(\hat{r}) + 2 \hat{r}^2 F''(\hat{r}) + \hat{r}^3 F'''(\hat{r}).
\end{equation}
This integrand is plotted in figure \ref{fig_app_gamma_integrand}, from which it is apparent that the integral is dominated by contributions from the viscous range, i.e. $r < 13 \eta$. Without considering the details of the integration, the manipulation so far shows that $\gamma$ is (approximately) independent of $\Rey_\lambda$ (neglecting weak  $\Rey_\lambda$-effects on the cutoff scale), and its precise value is difficult to determine because it will depend heavily on the details of the blending function used.

The integral can be written fully as
\begin{eqnarray}
I = \int\limits_{0}^{\infty} \left[\astrut \frac{16}{81} \left( 1 + \hat{r}^2 \right)^{-4/3} + \frac{416}{81} \left( 1 + \hat{r}^2 \right)^{-7/3} - \frac{304}{27} \left( 1 + \hat{r}^2 \right)^{-10/3} \right. \nonumber\\
- \left. \frac{2080}{81} \left( 1 + \hat{r}^2 \right)^{-13/3} + \frac{2560}{81} \left( 1 + \hat{r}^2 \right)^{-16/3} \astrut\right]\frac{d\hat{r}}{\hat{r}}.
\end{eqnarray}
To integrate, add
\begin{equation}
\left( \frac{16}{81} + \frac{416}{81} - \frac{304}{27} - \frac{2080}{81} + \frac{2560}{81} \right) \frac{\left(1 + \hat{r}^2\right)^{-1/3}}{\hat{r}} = 0,
\end{equation}
to the integrand and use the change of variables,
\begin{equation}
\zeta = 1 + \hat{r}^2, \hspace{0.05\linewidth} \frac{d\zeta}{2\left( \zeta - 1 \right)} = \frac{d\hat{r}}{\hat{r}}.
\end{equation}
As a result, the integral becomes,
\begin{eqnarray}
I = \frac{1}{2} \int\limits_{1}^{\infty} \left[\astrut -\frac{16}{81} \zeta^{-4/3} - \frac{416}{81}\zeta^{-7/3} \left(\zeta + 1\right) + \frac{304}{27} \zeta^{-10/3} \left( \zeta^2 + \zeta + 1 \right) \right. \nonumber\\
+ \left. \frac{2080}{81}\zeta^{-13/3} \left( \zeta^3 + \zeta^2 + \zeta + 1 \right) - \frac{2560}{81} \zeta^{-16/3}\left( \zeta^4 + \zeta^3 + \zeta^2 + \zeta + 1 \right) \astrut \right] d\zeta.
\end{eqnarray}
Then algebraic simplification
\begin{equation}
I = \frac{1}{2} \int\limits_{1}^{\infty} \left[ \frac{16}{81} \zeta^{-7/3} + \frac{432}{81} \zeta^{-10/3} - \frac{480}{81} \zeta^{-13/3} - \frac{2560}{81} \zeta^{-16/3}  \right] d\zeta,
\end{equation}
and completing the power-law integrations results in
\begin{equation}
I = -\frac{302}{91}.
\end{equation}
Substitution of this results leads to
\begin{equation}
\gamma = \frac{86}{1365} \approx 0.063.
\end{equation}

\section{Gaussian Fields Approximation for the Conditional Hessian of the Velocity Gradient\label{app_gaussian}}

This appendix details the derivation of  \eqref{eq_MODEL_GaussianViscousHessian} in the main text, following the method outlined in \citet{Wilczek2014}. The characteristic function of the turbulent velocity field,
\begin{equation}
\phi^{u}\left[\mathbf{\lambda}(\mathbf{x})\right] = \left\langle \exp\left( i \int \lambda_i(\mathbf{x}) u_i(\mathbf{x}) d\mathbf{x}\right) \right\rangle,
\end{equation}
contains all the statistical information necessary to compute the desired conditional mean, namely $\left\langle \left. \frac{\partial^2 A_{ij}}{\partial x_p \partial x_q} \right| \mathbf{\mathcal{A}} \right\rangle$. To make progress analytically, the turbulent velocity field is taken to be Gaussian, meaning that all $n$-point pdfs are joint-Gaussian,
\begin{equation}
\phi^{u}\left[\mathbf{\lambda}(\mathbf{x})\right] = \exp\left( -\frac{1}{2} \int \int \lambda_i(\mathbf{x}) B_{ij}(\mathbf{x},\mathbf{x}') \lambda_j(\mathbf{x}') d\mathbf{x} d\mathbf{x}' \right),
\end{equation}
where $B_{ij}$ is the two-point covariance tensor, which for homogeneous isotropic turbulence depends only on the separation vector, $\mathbf{r} = \mathbf{x} - \mathbf{x}'$, and has the form
\begin{equation}
B_{ij}(\mathbf{x},\mathbf{x}') = B_{ij}(\mathbf{r}) = \langle u_1^2 \rangle \left[ f(r) \delta_{ij} + \frac{1}{2} r f'(r) \left( \delta_{ij} - \hat{r}_i \hat{r}_j \right) \right],
\end{equation}
where $r = |\mathbf{r}|$ and $\hat{r}_i$ = $\frac{r_i}{r}$.
In this way the characteristic functional, when assumed Gaussian for isotropic turbulence, is uniquely specified by the longitudinal velocity correlation function,
\begin{equation}
f(r) = \frac{\left\langle u_1(\mathbf{x}) u_1(\mathbf{x}+r\mathbf{e}_1) \right\rangle}{\langle u_1^2 \rangle}.
\end{equation}
With integration by parts, the relationship between the characteristic functional for the velocity field and that of the velocity gradient field can be shown to be
\begin{equation}
\phi^A\left[ \mathbf{\Lambda} \right] = \phi^u\left[ -\nabla \cdot \mathbf{\Lambda} \right].
\end{equation}
Again, with integration by parts, substituting this relationship into the Gaussian characteristic functional for the velocity field,
\begin{equation}
\phi^A\left[\mathbf{\Lambda}(\mathbf{x})\right] = \exp\left( -\frac{1}{2} \int \int \Lambda_{ik}(\mathbf{x}) C_{ijk\ell}(\mathbf{x},\mathbf{x}') \Lambda_{j\ell}(\mathbf{x}') d\mathbf{x} d\mathbf{x}' \right),
\end{equation}
where
\begin{equation}
C_{ijk\ell}(\mathbf{x},\mathbf{x}') = \frac{\partial^2 B_{ij}}{\partial x_k \partial x_\ell'}(\mathbf{x},\mathbf{x}') = \left\langle A_{ik}(\mathbf{x}) A_{j\ell}(\mathbf{x}') \right\rangle,
\end{equation}
is the covariance tensor for the velocity gradient, which only depends on $\mathbf{r} = \mathbf{x} - \mathbf{x}'$. It is computed from the Hessian of the velocity covariance tensor,
\begin{eqnarray}
C_{ijk\ell}(\mathbf{r}) & = & - \frac{\partial^2 B_{ij}}{\partial r_k \partial r_\ell} = \langle u_1^2 \rangle \left[\astrut \left( -\frac{3}{2} \frac{f'(r)}{r} - \frac{1}{2} f''(r) \right) \left( \delta_{ij} \delta_{k\ell} \right)  + \left( \frac{1}{2} \frac{f'(r)}{r} \right) \left( \delta_{ik}\delta_{j\ell} + \delta_{i\ell} \delta_{jk} \right) \right. \nonumber\\
&& + \left( \frac{3}{2} \frac{f'(r)}{r} - \frac{3}{2} f''(r) - \frac{1}{2} r f'''(r) \right) \left( \delta_{ij} \hat{r}_k \hat{r}_\ell \right) + \nonumber\\
&& + \left( - \frac{1}{2} \frac{f'(r)}{r} + \frac{1}{2} f''(r) \right) \left( \delta_{i\ell} \hat{r}_j \hat{r}_k + \delta_{k\ell} \hat{r}_i \hat{r}_j + \delta_{j\ell} \hat{r}_i \hat{r}_k + \delta_{ik} \hat{r}_j \hat{r}_\ell + \delta_{jk} \hat{r}_i \hat{r}_\ell \right)\nonumber\\
&& \left.  \left( \frac{3}{2} \frac{f'(r)}{r} - \frac{3}{2} f''(r) + \frac{1}{2} r f'''(r) \right) \left( \hat{r}_i \hat{r}_j \hat{r}_k \hat{r}_\ell \right) \astrut\right].
\label{eq_app_Gaussian_C}
\end{eqnarray}
The desired statistical quantity in this exercise is
\begin{equation}
\nu \left\langle \left. \frac{\partial^2 A_{ij}}{\partial x_k \partial x_\ell} \right| \mathbf{\mathcal{A}} \right\rangle = \nu \lim\limits_{r\rightarrow 0} \frac{\partial^2}{\partial r_k \partial r_\ell} \left\langle A_{ij}(\mathbf{x}+\mathbf{r}) | \mathbf{\mathcal{A}}(\mathbf{x}) \right\rangle.
\end{equation}
Following exactly the steps outlined in Appendix B2 of \citet{Wilczek2014},
\begin{equation}
\left\langle A_{ij}(\mathbf{x}+\mathbf{r}) | \mathbf{\mathcal{A}}(\mathbf{x}) \right\rangle = C_{ikj\ell}(\mathbf{r}) C_{km\ell n}^{-1}(\mathbf{0}) \mathcal{A}_{mn},
\end{equation}
where equality at the origin means
\begin{equation}
C_{km\ell n}^{-1}(\mathbf{0}) = \frac{2}{15 \langle u_1^2 \rangle f''(0)} \left( -4 \delta_{km} \delta_{\ell n} - \delta_{kn} \delta_{m\ell} \right),
\label{eq_app_Gaussian_inverseC}
\end{equation}
see Appendix B1 of \citet{Wilczek2014} for details. Combining expressions,
\begin{equation}
\nu \left\langle \left. \frac{\partial^2 A_{ij}}{\partial x_k \partial x_\ell} \right| \mathbf{\mathcal{A}} \right\rangle = \nu \lim\limits_{r\rightarrow 0} \frac{\partial^2 C_{ikj\ell}}{\partial r_p \partial r_q} C_{km\ell n}^{-1}(\mathbf{0}) \mathcal{A}_{mn}.
\label{eq_app_Gaussian_evaluate}
\end{equation}
A tedious calculation by twice taking the gradient of  \eqref{eq_app_Gaussian_C} results in,
\begin{eqnarray}
\lim\limits_{r\rightarrow 0} \frac{\partial^2 C_{ikj\ell}}{\partial r_p \partial r_q} = \langle u_1^2 \rangle f^{(4)}(0) \left[ - \left( \delta_{ik}\delta_{j\ell}\delta_{pq} + \delta_{ik}\delta_{jp}\delta_{\ell q} + \delta_{ik}\delta_{jq}\delta_{\ell p} \right)
\right. \nonumber \\
+ \tfrac{1}{6} \left( \delta_{ij}\delta_{k\ell}\delta_{pq} + \delta_{i\ell} \delta_{kj} \delta_{pq} + \delta_{i\ell}\delta_{kp} \delta_{jq}   + \delta_{i\ell}\delta_{jp}\delta_{kq} + \delta_{ip} \delta_{k\ell} \delta_{jq} + \delta_{iq} \delta_{k\ell} \delta_{jp} \right. \nonumber\\
\left.\left. + \delta_{j\ell}\delta_{kp}\delta_{iq} + \delta_{j\ell} \delta_{ip} \delta_{kq} + \delta_{ij} \delta_{kp} \delta_{\ell q} + \delta_{ij} \delta_{\ell p} \delta_{kq} + \delta_{kj} \delta_{ip} \delta_{\ell q} + \delta_{kj}\delta_{\ell p} \delta_{iq} \right)
\right]
\label{eq_app_gaussian_HessianC}
\end{eqnarray}
Substitution of  \eqref{eq_app_Gaussian_inverseC} and \eqref{eq_app_gaussian_HessianC} into  \eqref{eq_app_Gaussian_evaluate}, followed by a tedious calculation of tensor contractions yields,
\begin{eqnarray}
\nu \left\langle \left. \frac{\partial^2 A_{ij}}{\partial x_k \partial x_\ell} \right| \mathbf{\mathcal{A}} \right\rangle =  \frac{2 \nu f^{(4)}(0)}{15 f''(0)} \left[ \left( \frac{23}{6} \mathcal{A}_{ij} + \frac{1}{3} \mathcal{A}_{ji} \right)\delta_{pq} + \left( \frac{23}{6} \mathcal{A}_{iq} + \frac{1}{3} \mathcal{A}_{qi} \right)\delta_{jp} \right. \nonumber\\
+ \left( \frac{23}{6} \mathcal{A}_{ip} + \frac{1}{3} \mathcal{A}_{pi} \right)\delta_{jq} - \left( \frac{5}{6} \mathcal{A}_{jq} + \frac{5}{6} \mathcal{A}_{qj} \right) \delta_{ip} \nonumber\\
\left. - \left( \frac{5}{6} \mathcal{A}_{jp} + \frac{5}{6} \mathcal{A}_{pj} \right) \delta_{iq} - \left( \frac{5}{6} \mathcal{A}_{pq} + \frac{5}{6} \mathcal{A}_{qp} \right) \delta_{ij}  \right],
\end{eqnarray}
which can be written in the form of  \eqref{eq_MODEL_GaussianViscousHessian} with  \eqref{eq_MODEL_GaussianViscousHessian_tensors}.

\section{Determination of $\delta$ Using the Enstrophy Balance\label{app_delta}}

Using the result of Appendix \ref{app_gaussian}, the back-in-time velocity gradient Hessian is given by
\begin{equation}
\nu \left\langle \left. \frac{\partial^2 A_{ij}}{\partial X_p \partial X_q} \right| \mathcal{A} \right\rangle = \delta \left( T_{ij} \delta_{pq} + T_{iq} \delta_{jp} + T_{ip} \delta_{jq} - \frac{2}{21} S_{jq} \delta_{ip} - \frac{2}{21} S_{jp} \delta_{iq} - \frac{2}{21} S_{pq} \delta_{ij} \right),
\end{equation}
where the coefficient $\delta$ can be written in terms of the enstrophy dissipation,
\begin{equation}
\delta = \nu \frac{7}{3} \frac{f^{(4)}(0)}{f''(0)} = - \tau_\eta^2 \nu \left\langle \frac{\partial \omega_i}{\partial X_j} \frac{\partial \omega_i}{\partial X_j}\right\rangle.
\end{equation}
Note that since the Gaussian fields evaluation is back-in-time, so this can be interpreted as the back-in-time enstrophy dissipation. By definition, the RFD-style mapping used to generate the approximate back-in-time values keeps velocity gradients constant, but not velocity Hessians. Therefore, the enstrophy production $\langle \omega_{i} S_{ij} \omega_{j}\rangle$ is constant under the mapping but the enstrophy dissipation is not constant. Two choices are thus available: apply the enstrophy balance for the back-in-time enstrophy dissipation, or try to invert the mapping effect on the enstrophy dissipation to apply the balance at the present time. It is the opinion of the authors that the second option is desirable, since it leads to the application of the enstrophy balance at the present time rather than back-in-time.

Thus, by modeling choice, the relevant enstrophy balance is
\begin{equation}
\left\langle \omega_i S_{ij} \omega_j \right\rangle = \nu \left\langle \frac{\partial \omega_i}{\partial x_j} \frac{\partial \omega_i}{\partial x_j}\right\rangle.
\end{equation}
To map the enstrophy dissipation forward in time,
\begin{equation}
\left\langle \frac{\partial \omega_i}{\partial X_j} \frac{\partial \omega_i}{\partial X_j}\right\rangle = \left\langle \frac{\partial x_k}{\partial X_j} \frac{\partial \omega_i}{\partial x_k} \frac{\partial \omega_i}{\partial x_\ell} \frac{\partial x_\ell}{\partial X_j} \right\rangle = \left\langle C_{k\ell} \frac{\partial \omega_i}{\partial x_k} \frac{\partial \omega_i}{\partial x_\ell} \right\rangle \approx C_{k\ell} \left\langle \frac{\partial \omega_i}{\partial x_k} \frac{\partial \omega_i}{\partial x_\ell} \right\rangle.
\end{equation}
In the last step, the value of $C_{k\ell}$ is localized by approximation, so that no ensemble averages are needed to advance the model stochastic equations. Finally, the enstrophy dissipation tensor is assumed isotropic,
\begin{equation}
\left\langle \frac{\partial \omega_i}{\partial x_k} \frac{\partial \omega_i}{\partial x_\ell} \right\rangle \approx \frac{1}{3} \left\langle \frac{\partial \omega_i}{\partial x_j} \frac{\partial \omega_i}{\partial x_j} \right\rangle \delta_{k\ell}.
\end{equation}
Substituting, the resulting enstrophy balance is
\begin{equation}
\left\langle \omega_i S_{ij} \omega_j \right\rangle = \frac{3\nu}{C_{kk}} \left\langle \frac{\partial \omega_i}{\partial X_j} \frac{\partial \omega_i}{\partial X_j} \right\rangle.
\end{equation}
Using the isotropic relation $\langle \omega_i S_{ij} \omega_j \rangle = - \tfrac{7\mathcal{S}}{6\sqrt{15}\tau_\eta^3}$ on the left side and the definition of $\delta$ in terms of enstrophy dissipation on the right side, the result is
\begin{equation}
\delta = \frac{C_{kk}}{3} \frac{7}{6\sqrt{15}} \frac{\mathcal{S}}{\tau_\eta}.
\end{equation}
The result given by \citet{Wilczek2014} is recovered when the mapping is removed, $D_{ij} = \delta_{ij}$, so that $C_{kk} = 3$. In this way, the $\delta$ coefficient itself depends on the recent deformation. This provides the convenience of an additional non-linearity in the viscous term to prevent unwanted singularities while advancing the stochastic differential equation.

As a final note, the scaling of $\delta \sim \tau_\eta^{-1}$ contradicts the RFD model for the viscous Laplacian, which used the integral timescale and thus introduced a $\Rey_{\lambda}^{-1}$ scaling for the viscous term. While $\Rey_\lambda$ dependence can be introduced in the present model through the skewness coefficient, the similar difficulties as encountered by the RFD model are seen when going to large Reynolds numbers. It is the authors' view that a fixed skewness coefficient, $\mathcal{S} = -0.6$, is appropriate for the present model's the level of fidelity.

\bibliographystyle{jfm_alt}
\bibliography{LaVelGrad.bib}

\end{document}